\documentclass{jfm}
\usepackage{graphicx}
\usepackage{epstopdf, epsfig}

\usepackage[utf8]{inputenc}
\usepackage{amsmath,bm}
\usepackage{float}
\usepackage{amssymb}
\usepackage{xcolor}
\usepackage{subfigure}
\usepackage{amsfonts}
\usepackage{comment}

\title{Settling of two rigidly connected spheres}

\author{Z. Maches\aff{1}, M. Houssais\aff{2}, A. Sauret\aff{1} \and E. Meiburg\aff{1}\corresp{\email{meiburg@engineering.ucsb.edu}}}

\affiliation{\aff{1}Department of Mechanical Engineering, University of California at Santa Barbara, Santa Barbara, CA 93106, USA
\aff{2}Department of Physics, Clark University, Worcester, MA 01610, USA}

\begin{document}

\maketitle

\begin{abstract}
Laboratory experiments and particle-resolved simulations are employed to investigate the settling dynamics of a pair of rigidly connected spherical particles of unequal size. They yield a detailed picture of the transient evolution and the terminal values of the aggregate's orientation angle and its settling and drift velocities as functions of the aspect ratio and the Galileo number $Ga$, which denotes the ratio of buoyancy and viscous forces acting on the aggregate. At low to moderate values of $Ga$, the aggregate's orientation and velocity converge to their terminal values monotonically, whereas for higher $Ga$-values the aggregate tends to undergo a more complex motion. If the aggregate assumes an asymmetric terminal orientation, it displays a nonzero terminal drift velocity. For diameter ratios much larger than one and small $Ga$, the terminal orientation of the aggregate becomes approximately vertical, whereas when $Ga$ is sufficiently large for flow separation to occur, the aggregate orients itself such that the smaller sphere is located at the separation line. Empirical scaling laws are obtained for the terminal settling velocity and orientation angle as functions of the aspect ratio and $Ga$ for diameter ratios from 1 to 4 and particle to fluid density ratios from 1.3 to 5. An analysis of the accompanying flow field shows the formation of vortical structures exhibiting complex topologies in the aggregate's wake, and it indicates the formation of a horizontal pressure gradient across the larger sphere, which represents the main reason for the emergence of the drift velocity.
\end{abstract}


%%%%%%%%%%%%%%%%%%%%%%%%%%%%%%%%%%%%%%%%%%%%%%%%%%%%%%
%%%%%%%%%%%%%%%%%%  Introduction  %%%%%%%%%%%%%%%%%%%%
%%%%%%%%%%%%%%%%%%%%%%%%%%%%%%%%%%%%%%%%%%%%%%%%%%%%%%

\section{Introduction}\label{sec:intro}
The settling dynamics of irregularly shaped particles and particle aggregates are of considerable interest for a variety of environmental transport processes and engineering applications. Examples include the settling of mud, clay and silt \citep[][]{WINTERWERP20021339, 10.2110/jsr.2011.62, te2015hindered}, as well as marine snow \citep[][]{lo19883330339, DIERCKS1997385} and microplastics \citep[][]{KHATMULLINA2017871, Wang2021, Yan2021}. Similar issues arise with regard to the dynamics of ice particles in clouds \citep[][]{PhysRevLett.119.254501} and in the context of industrial processes involving particle flocculation \citep[][]{LICSKO1997103, KURNIAWAN200683}. Environmental systems and engineering applications such as deep-sea mining \citep[]{meiburg2010turbidity,peacock2018deep,gillard2019physical,ouillon_kakoutas_meiburg_peacock_2021,wells2021turbulence} may also be strongly affected by where particles are transported and deposited, so that it is desirable to have accurate predictions of their settling and drift velocities as functions of their geometry.

While the settling of individual spheres across different flow regimes is a well-studied topic that has received much attention \citep[][]{leal1980particle, WR018i006p01615, doi:10.1061/(ASCE)0733-9372(2003)129:3(222), YANG2015219}, more recent studies have addressed the settling dynamics of irregularly shaped objects in inertial flows, such as oblate spheroids \citep[][]{MORICHE2021103519}, cubes and tetrahedra \citep[][]{doi:10.1063/1.4892840}, slender bodies \citep[][]{khayat_cox_1989}, and disks \citep[][]{heisinger_newton_kanso_2014}. These studies show that the spatial orientation and settling velocity of a body are highly dependent on its geometry. Changes in the size and shape of an aggregate can greatly affect its drag \citep{LOTH2008342,li2020settling}, leading to large changes in the settling rate.

An additional key feature of aggregate shape concerns the effective porosity and permeability. 
The formation of particle aggregates can be triggered by a variety of forces, such as van der Waals forces \citep[][]{VISSER19891} or biocohesive forces caused by the bonds between organic molecules \citep[][]{biological_coh}. Recent numerical works have taken first steps towards simulating such cohesive forces, as analyzed in the investigation of aggregates with brittle tensile bonds by \citep[][]{https://doi.org/10.1002/2014JF003330}, or in the study of the settling dynamics of cohesive sediment held together by van der Waals forces \citep[][]{vowinckel_withers_luzzatto-fegiz_meiburg_2019}. These aggregates, in turn, due to their porosity, allow for fluid to pass through the pore spaces, which influences the settling behavior \citep{PRAIRIE201528}. The size and distribution of these pore spaces are controlled by the geometry of the aggregate, such that the arrangement of particles leads to aggregates with highly variable settling velocities.

In contrast to bodies with fore-aft asymmetry, for nonsymmetric bodies the settling motion can become significantly more complex, causing them to settle along complicated trajectories \citep{doi:10.1021/es950604g,TANG2002210}. In the present work, we consider the sedimentation of rigid bodies consisting of two connected spheres of unequal diameters. Except for a simplified model variant \citep{candelier_mehlig_2016}, the dynamics of this case has not yet received much attention, while the similar situation of particle pairs with equal diameters but unequal densities has been examined in some depth by \citet{nissanka_ma_burton_2023}. The existing literature on the topic primarily focuses of the the case of highly viscous flows at or just above the Stokes limit. As we will see, the asymmetric particle pair represents a relatively simple case that can exhibit complicated settling dynamics depending on the ratio of the diameters.

The present study focuses on the dynamics of elementary aggregates consisting of two spherical particles of different sizes. We employ both particle-resolved simulations as well as laboratory experiments, in order to investigate the dynamics of such aggregates settling in a fluid at rest. The problem configuration is described in detail in Section \ref{sec:methods}, including the experimental and numerical approaches, as well as validation results. Section \ref{sec:num_only} presents comparisons of experimental and simulation results, and it explores the influence of the ratios of the diameters and also of the gravitational and viscous forces in depth. Scaling laws are obtained for the dynamics of the aggregate as a function of these dimensionless parameters, and the accompanying fluid flow field is analyzed. Section \ref{sec:conclusion} summarizes the main findings of the study, and discusses potential further extensions.

%%%%%%%%%%%%%%%%%%%%%%%%%%%%%%%%%%%%%%%%%%%%%%%%%%%%%%
%%%%%%%%%%%%%%%%%%%%%  Methods  %%%%%%%%%%%%%%%%%%%%%%
%%%%%%%%%%%%%%%%%%%%%%%%%%%%%%%%%%%%%%%%%%%%%%%%%%%%%%

\section{Methods}\label{sec:methods}

\subsection{Problem definition}\label{sec:phenomeno}

\begin{figure}
\centering
  \includegraphics[width=\textwidth]{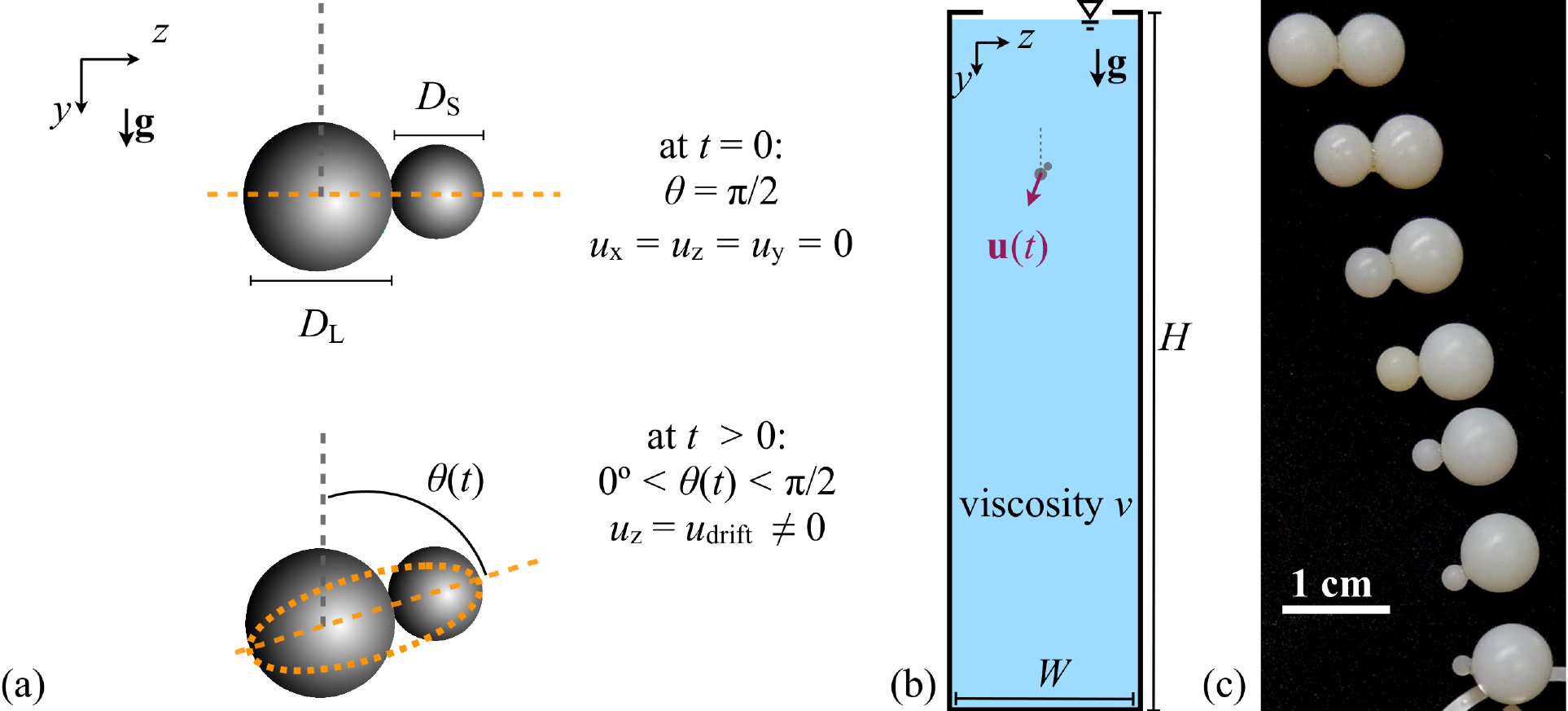}
  \caption{Schematics (a) of the pair of particles considered at $t = 0$ (top) and at a later time (bottom), (b) of the experimental setup, and (c) of example aggregates prepared in the laboratory for various values of $\alpha$. Straight orange dashed lines in (a) indicate the lines forming the orientation angle $\theta$, and the curved orange dashed lines indicate the elliptical shape employed to extract the orientation angle from the experimental data during image processing.}
\label{fig:Figure1}
\end{figure}

We aim to characterize the dynamics of a pair of rigidly connected spherical particles with density $\rho_{\rm p}$ that settles in a fluid of density $\rho_{\rm f}<\rho_{\rm p}$ and kinematic viscosity $\nu$, in a large tank of square cross-section with width $W$ and height $H$, as illustrated in figure \ref{fig:Figure1}. The large and small particles have a diameter $D_{\rm L}$ and $D_{\rm S}$, respectively with $D_{\rm L} \ge D_{\rm S}$, so that the geometry of the aggregate is characterized by the diameter size ratio
\begin{equation}
    \alpha = D_{\rm L}/D_{\rm S} \ .
\end{equation}

At the initial time, $t=0$, the particle centers of gravity are aligned horizontally in the $z$-direction, and both the fluid and the particles are at rest, as shown in figure \ref{fig:Figure1}(a). Gravity points in the $y$-direction. Due to the symmetry of the particle pair and the moderate influence of fluid inertia considered in the present study, the motion of the aggregate in the $x$-direction is negligible compared to that in the ($y,z$)-plane. We hence describe the motion of the particle pair in terms of the vertical settling and the lateral drift velocities of its center of mass, $u_y$ and $u_z$, respectively, as well as its orientation angle $\theta$, defined as the angle between the upward vertical axis and the line connecting the spheres' centers.

\subsection{Experimental methods}\label{sec:exp}

\begin{table}
  \begin{center}
  \def~{\hphantom{0}}
  \begin{tabular}{c|cccccc}
 &$\rho '$ &$Ga$ & $\alpha$ & $H$ & $W$ & $\theta(t=0)$    \\ \hline
 Exp.&1.34, 1.31&8-22&1-7&59.4 cm&14.1 cm&$\pi/2$ \\
 Num. (Exp.) &1.31&8-22 &1-2.4& $50D_{\rm L}$&$10D_{\rm L}$&$\theta_1$\\
 Num. ($Ga \leq 13$)&1.31&4-13 &1-2& $50D_{\rm L}$&$10D_{\rm L}$&$\pi/2$\\
 Num. ($18\leq Ga\leq 42$) &1.31&18-12 &1-4& $80D_{\rm L}$&$10D_{\rm L}$&$\pi/2$\\
 Num. ($Ga = 75$) &1.31&75 &1-4& $120D_{\rm L}$&$10D_{\rm L}$&$\pi/2$
  \end{tabular}
  \caption{Key parameter ranges considered in the experiments and numerical simulations performed in the present study, for the diameter aspect ratio $\alpha$, density ratio $\rho'$, Galileo number $Ga$, dimensional domain height $H$ and width $W$, and initial orientation angle $\theta(t=0)$. For the numerical simulations corresponding to the experimental cases, we define $\theta_1$ to be the first recorded value of the orientation angle obtained via measurement.}
  \label{tab:params}
  \end{center}
\end{table}

The experiments were conducted in a tall tank of height $H = 59.4$ cm and square cross-section of dimensions $W \times W=14.1 \times 14.1\,{\rm cm^2}$. The tank was filled with one of two types of mineral oils: STE Food Grade 200 (dynamic viscosity $\mu_{\rm f,1} = 96\,{\rm mPa \, s}$ and density $\rho_{\rm f,2} = 864 \,{\rm kg \, m^{-3}}$ at 20$^o$C), or STE Food Grade 70 (dynamic viscosity $\mu_{\rm f,2} = 23\,{\rm mPa \, s}$ and density $\rho_{\rm f,1}  = 850\,{\rm kg \, m^{-3}}$ at 20 $^o $C). 

Aggregates made of two spheres were released in a horizontally aligned configuration (see figure \ref{fig:Figure1}(a), top). The aggregates were manufactured by gluing together two nylon spheres of density  $\rho_{\rm p} = 1135\,{\rm kg \, m^{-3}}$ (figure \ref{fig:Figure1}(c)). The larger particle diameters were $D_{\rm L} = 4.76,\, 6.35,\, 9.53,\, {\rm and}\, 11.11\,{\rm mm}$, while the smaller particle size ranged from $D_{\rm S} = 1.59 \,{\rm mm}$ to $11.11 \,{\rm mm}$. This leads to values of $\alpha = D_{\rm L}/D_{\rm S}$ in the range $1 \leq \alpha \leq 7$, with the majority of the aggregates having $\alpha \leq 4$, which is the range in which the change in dynamics is most pronounced. For the sake of experimental simplicity, some of the experiments at lower viscosity were performed in a smaller tank, without changing quantitatively the results. The parameter ranges for both the experiments and the numerical simulations are indicated in table \ref{tab:params}.

The aggregate was first immersed in a beaker to be coated with the same oil as in the bath to prevent bubble entrainment during its later immersion into the tank. Then, the aggregate was gently submerged in the tank about 1 cm below the fluid surface with an adjustable metallic wrench. The wrench rested on a flat horizontal support placed on the top of the tank, and was opened manually to release the aggregate. The experimental setup was backlit with a LED panel and a Nikon camera recorded the settling of the aggregate in the fluid at 60 frames per second (see video examples in Supplementary Materials), so that the instantaneous velocity and angle could be extracted.
Three copies of each aggregate geometry ($D_{\rm L}$, $\alpha$) were built from three distinct pairs of spheres, and each aggregate's settling motion was recorded to later average the results over the effective variability in sizes and densities of the nylon spheres. 

In most cases, the aggregates reached a quasi-steady state at the end of an experiment, \textit{i.e.}, the aggregate velocity and orientation angle no longer varied with time. Only a few cases did not reach a steady state, due to the limited size of the experimental domain. $\theta$ was determined by aligning an elliptical shape onto the images of the settling pair, and calculating the angle formed by its major axis with the vertical direction (see the curved dashed lines in figure \ref{fig:Figure1}(a)).

A few experimental cases, where the aggregate rotated significantly out of the ($y,z$)-plane, were detected via tracking the apparent aggregate shape, and subsequently discarded. The influence of fluid inertia in the present work is not large enough to expect out-of-plane rotation, so that we attribute any rotation in this direction in the experiments to inadvertent angular velocity generated during aggregate release. The numerical simulations support this approach by showing negligible out-of-plane rotation for all cases.

For detailed comparisons with the numerical simulations, a series of settling experiments of single spheres (the same constituting the aggregates) were performed in the same fluids and tanks as the aggregate experiments. The viscosity values at the average room temperature ($20^{\rm o}{\rm C}$) for the two fluids, $\mu_{\rm f,1}(T = 20^{\rm o}{\rm C})$ and $\mu_{\rm f,2}(T = 20^{\rm o}{\rm C} )$,  were obtained by comparing the experimental terminal settling velocities, $u_{\rm term}$, with a classical drag coefficient law \citep[][]{schiller1933uber}
\begin{equation}
C_D = \frac{8 F_{\rm drag}}{\rho_{\rm f} \, u_{\rm term}^2 \pi D_{\rm L}^2} = \frac{8 \left(m_{\rm p}-m_{\rm f}\right) g}{\rho_{\rm f} \, u_{\rm term}^2 \pi D_{\rm L}^2} = \frac{24}{\Rey}\left(1 + 0.15\Rey^{0.687}\right),\label{eq:schill}
\end{equation}
where the Reynolds number is defined as
\begin{equation}\label{eq:re}
    \Rey = \frac{D_{\rm L} \, u_{\rm term}}{\nu},
\end{equation}
with the single (or large) particle diameter $D_{\rm L}$. $F_{\rm drag}$ denotes the drag force on the spherical particle, $m_{\rm p}$ its mass, $m_{\rm f}$ the mass of fluid in a volume the size of the particle, and $\nu = \mu_{\rm f}/\rho_{\rm f}$ is the kinematic viscosity. 

In order to build a sufficiently large tank without being able to determine the time required to reach a terminal (steady-state) velocity, we took the following approach. To decide on the tank dimensions, oil viscosities, and particle sizes, we chose a target range of Reynolds numbers $\Rey \in \left[1, 100\right]$, and tank dimensions such that the time to settle over the distance of the tank height $H$ is greater than $20 \times  D_{\rm max}/u_{\rm term}(D_{\rm max})$ and the tank width is such that $W> 10 \times D_{\rm max}$, with $D_{\rm max} = 11.1\;{\rm mm}$. We determined the corresponding range of $u_{\rm term}(D_{\rm max})$ using equations (\ref{eq:schill}) and (\ref{eq:re}). The experimental data presented here are those for which $H$ was large enough so that the settling velocity reached its steady-state.

To account for slight temperature variations in the room (between $17^{\rm o}{\rm C}$ and $23^{\rm o}{\rm C}$), we measured the temperature dependence of the oil viscosities with a plate-plate geometry on an Anton Paar MCR 302 rheometer. These measurements provided $\mu_{{\rm f},i,{\rm rheom}}(T)$ for each of the two oils.
We then fitted the measured data (provided in Supplementary Materials) by the curves $\mu_{{\rm f},i}(T) = c_{i} \mu_{{\rm f}, i,{\rm rheom}}(T)$, with $c_{\rm 1} = 1.15$ and $c_{\rm 2} = 1.1$, so that the rheometer measurements agree with our best determination of absolute viscosity values from the settling sphere experiments, obeying equation (\ref{eq:schill}) at $20^{\rm o}{\rm C}$. To summarize, we base the absolute viscosity value on the comparison with equation (\ref{eq:schill}), and the relative variation of the viscosity with temperature on the rheometer measurements.

\subsection{Governing equations}
The fluid flow is governed by the incompressible continuity equation
\begin{align}
    \nabla \cdot \bm{u} &= 0,
\end{align}
and the unsteady Navier-Stokes equation
\begin{align}
    \frac{\partial \bm{u}}{\partial t} + \left(\bm{u}\cdot \nabla \right)\bm{u} &= -\frac{1}{\rho_{\rm f}}\nabla p +\nu\nabla^2\bm{u} +\frac{1}{\rho_{\rm f}}\bm{f}_{\rm IBM} \ ,
\end{align}
where $\bm{u}$ represents the fluid velocity, $t$ the time, $p$ the pressure with the hydrostatic component subtracted out, and $\bm{f}_{\rm IBM}$ the distributed force exerted on the fluid by the particles, which is calculated based on an Immersed Boundary Method (IBM) approach, as will be discussed below.

The motion of the finite-size solid particles is governed by
\begin{align}
    m_i\frac{{\rm d}\bm{u}_i}{{\rm d}t} &= \oint\limits_{\Gamma_i}\bm{\tau}\cdot\bm{n}{\rm d}A+V_i\left(\rho_{{\rm p}, i}-\rho_{\rm f}\right)\bm{g}+\bm{F}_{{\rm b},i} \ , \\
    I_i\frac{{\rm d}\bm{\bm{\omega}}_i}{{\rm d}t} &= \oint\limits_{\Gamma_i}\bm{r}\times \left(\bm{\tau}\cdot\bm{n}\right){\rm d}A+\bm{M}_{{\rm b},i} \ ,
\end{align}
where $m_i$ indicates the mass of the $i$-th particle, $\bm{u}_i$ and $\bm{\bm{\omega}}_i$ its translational and angular velocities, respectively, $V_i$ its volume, $\Gamma_i$ its surface, $\rho_{{\rm p}, i}$ its density, and $I_i$ its moment of inertia. $\bm{F}_{{\rm b},i}$ and $\bm{M}_{{\rm b},i}$ represent the sum of all forces and moments acting on particle $i$ via rigid bonds with other particles, $\bm{g}$ denotes the gravity vector, and $\bm{\tau}$ indicates the hydrodynamic stress tensor.

We also define $\bm{n}$ to be the outward normal vector on $\Gamma_i$, and $\bm{r} = \bm{x}-\bm{x}_i$ to be the position vector from the particle center $\bm{x}_i$ to the surface point $\bm{x}$. Unless otherwise stated, we restrict our study to the case of two particles with varying diameters and identical densities, so that $\rho_{{\rm p}, i} = \rho_{\rm p}$.

\subsection{Nondimensionalization}

We define characteristic length, velocity, time and pressure scales as
\begin{align} 
    x_{\rm ref} = D_{\rm  L} \ \ \ , \ \ \ u_{\rm ref} = \sqrt{g' \,D_{\rm L}} \ \ \ , \ \ \ t_{\rm ref} = \sqrt{D_{\rm L} / g'} \ \ \ , \ \ \ p_{\rm ref} = \rho_{\rm f} \, u_{\rm ref}^2\ , \label{eqn:ref}
\end{align}
where $g'=g (\rho ' - 1)$ is the the reduced gravity, with $\rho '=\rho_{\rm p}/\rho_{\rm  f}$ denoting the ratio of particle to fluid density. We remark that we choose the diameter of the larger sphere as our characteristic length scale, in order to focus on how the behavior of the aggregate deviates from that of a single spherical particle when an additional, smaller sphere is attached to it.
In this way, the governing dimensionless equations take the form
\begin{align}
    \nabla \cdot \hat{\bm{u}} &= 0 \ ,\\
    \frac{\partial \hat{\bm{u}}}{\partial t} + \left(\hat{\bm{u}}\cdot \nabla \right)\hat{\bm{u}} &= -\nabla\hat{p}+\frac{1}{Ga}\nabla^2\hat{\bm{u}} +\hat{\bm{f}}_{\rm IBM} \ , \\
   \frac{{\rm d}\hat{\bm{u}}_i}{{\rm d}\hat{t}} &= \frac{1}{\rho'}\oint\limits_{\hat{\Gamma}_i}\hat{\bm{\tau}}\cdot\bm{n}{\rm d}\hat{A}+\hat{\bm{F}}_{{\rm b},i} - \frac{1}{\rho'} \ , \\
    \frac{{\rm d}\hat{\bm{\omega}}_i}{{\rm d}\hat{t}} &= \frac{1}{\rho'} \oint\limits_{\hat{\Gamma}_i}\hat{\bm{r}}\times \left(\hat{\bm{\tau}}\cdot\bm{n}\right){\rm d}\hat{A}+\hat{\bm{M}}_{{\rm b},i} \ ,
\end{align}
where $\hat{\bm{u}}_i = \bm{u}_i/u_{\rm ref}$ and $\hat{\bm{\omega}}_i=\bm{\omega}_i \, t_{\rm ref}$ are the dimensionless particle translational and angular velocities, respectively. $\hat{\bm{F}}_{{\rm b},i} = \bm{F}_{{\rm b},i}/\left(m_ig'\right)$ and $\hat{\bm{M}}_{{\rm b},i} = D_{\rm L}\bm{M}_{{\rm b},i}/\left(I_ig'\right)$ denote the dimensionless bond force and moment. The Galileo number,
\begin{equation}\label{eq:re_term}
    Ga =\frac{ D_{\rm L}\, u_{\rm ref}}{\nu}  = \frac{D_{\rm L}\sqrt{D_{\rm L}g\left(\rho '-1\right)}}{\nu},\;
\end{equation}
indicates the ratio of gravitational to viscous forces. In the rest of the manuscript, we will discuss dimensionless quantities only, so that we omit the $\, \hat{} \,$ symbol.

\subsection{Numerical approach}\label{sec:num}
We solve the governing equations based on the Immersed Boundary Method (IBM) implementation in the code PARTIES, as described in \citet[][]{BIEGERT2017105} and \citet[][]{Bieg_dis}. In summary, it employs an Eulerian mesh with uniform spacing $h = \Delta x = \Delta y = \Delta z$ for calculating the fluid motion. The viscous and convective terms are discretized by second-order central differences, and a direct solver based on Fast Fourier Transforms is used to obtain the pressure. Time integration is generally performed by a third-order Runge-Kutta scheme, although viscous terms are integrated implicitly with second-order accuracy. To compute $\bm{f}_{\rm IBM}$, we implement the method of \citet[][]{KEMPE20123663}, which employs a mesh of Lagrangian marker points on the surface of each particle. At these marker points, we interpolate between the Eulerian and Lagrangian grids in order to determine the value of $\bm{f}_{\rm IBM}$ such as to ensure the no-slip and no-normal flow conditions at the particle surface. In the bond region we found that for the range of $Ga$ considered any overlap between marker points did not have a noticeable effect on the settling dynamics, and as such did not deactivate those close to the contact point as done in \citet{Bieg_dis}. Additional details regarding the implementation for individual spherical particles, along with validation results, can be found in the references by \citet[]{vowinckel_withers_luzzatto-fegiz_meiburg_2019} and \citet[][]{zhu_he_zhao_vowinckel_meiburg_2022}. 

Those earlier investigations did not consider rigid bonds between the individual particles, which are implemented here for the first time. Hence, we provide in the following a detailed description of the method. First, let us consider the particles $P_1$ and $P_2$, bonded at a shared contact point $\bm{x}_{\rm c}$. We define the contact plane between these particles as the plane that crosses through $\bm{x}_{\rm c}$ and is orthogonal to the line between the particle centers. The bond is implemented via a corrective force and moment designed to hold the particles together at the contact point. This will ensure that, at all times, $\dot{\bm{x}}_{\rm c,1} = \dot{\bm{x}}_{\rm c,2}$, \textit{i.e.}, the velocities of the contact point evaluated on the surface of each particle are the same, and $\bm{\omega}_{\rm 1}=\bm{\omega}_{\rm 2}$, \textit{i.e.}, the two particles have the same angular velocity so that they rotate as a solid object. The first condition prevents the particles from moving apart or from sliding in opposite directions, while the second prevents the particles from rolling along each other's surface, and from having different angular velocities around the axis connecting them.
\begin{figure}
    \centering
    \includegraphics[width=0.4\textwidth]{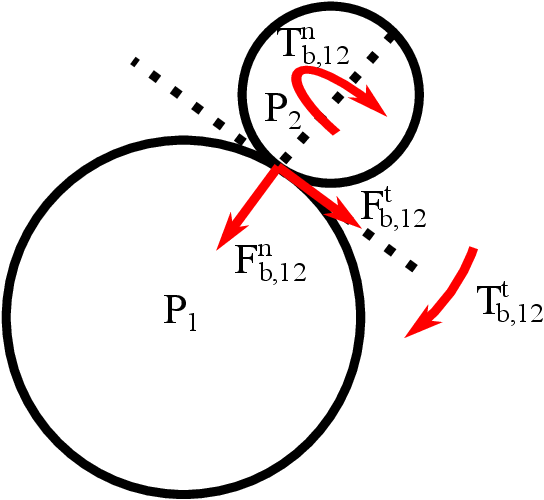}
    \caption{Sketch of the forces and torques acting on the two spherical particles $P_1$ and $P_2$ that form the aggregate, due to the rigid bond connecting them.}
    \label{fig:bond_sketch}
\end{figure}
We note that, in principle, the immersed boundary method for complex aggregates could be implemented without the use of bonds, by treating the aggregate as a single rigid object and distributing the Lagrangian marker points over its more complex surface. However, we found the present approach to be more straightforward to parallelize for aggregates of complex shapes, and also to be easily applicable to aggregates that may undergo breakup, which are scenarios that we plan to address in the future. We approximate the rigid bond through the use of spring forces and moments, as illustrated in figure \ref{fig:bond_sketch}, in order to satisfy an approximate form of the above conditions.

To compute the force and moment due to the bond, we define the bond force and moment acting on particle $i$ by particle $j$ by splitting each into their normal and tangential components
\begin{align}
    \bm{F}_{{\rm b},ij} &= \bm{F}_{{\rm b},ij}^n+\bm{F}_{{\rm b},ij}^t, \\
    \bm{M}_{{\rm b},ij} &= \bm{r}_{{\rm c}, i}\times \bm{F}_{{\rm b},ij}^t +\bm{M}_{{\rm b},ij}^n+\bm{M}_{{\rm b},ij}^t,
\end{align}
with $\bm{F}_{{\rm b},ij}^n$ and $\bm{F}_{{\rm b},ij}^t$ being the components of the bond force normal and tangential to the contact plane between the two particles (with the normal and tangential components marked in superscript as $n$ and $t$, respectively), $\bm{M}_{{\rm b},ij}^n$ and $\bm{M}_{{\rm b},ij}^t$ being the corresponding components of the moment, and $\bm{r}_{{\rm c}, i} = \bm{x}_{{\rm c}}-\bm{x}_{i}$ being the distance between the contact point and the center of the particle $i$.

We compute the normal and tangential forces and moments using a variation of the model described by \citet[][]{POTYONDY20041329}, which considers the particle-particle bond as an area of infinitely small springs along the contact plane. We initialize the force and moment for each bond to be zero at $t=0$. Then, at each time step, we compute the dimensionless forces and moments as
\begin{align}
    \bm{F}_{{\rm b},ij}^n\left(t+\Delta t\right) &= \bm{R}\bm{F}_{{\rm b},ij}^n\left(t\right)-k_{1}\Delta \dot{\bm{x}}_{{\rm c},ij}^n\Delta t,\\
    \bm{F}_{{\rm b},ij}^t\left(t+\Delta t\right) &= \bm{R}\bm{F}_{{\rm b},ij}^t\left(t\right)-k_{1}\Delta \dot{\bm{x}}_{{\rm c},ij}^t\Delta t,\\
    \bm{M}_{{\rm b},ij}^n\left(t+\Delta t\right) &= \bm{R}\bm{M}_{{\rm b},ij}^n\left(t\right)-k_{2}\Delta \bm{\omega}_{ij}^n\Delta t,\\
    \bm{M}_{{\rm b},ij}^t\left(t+\Delta t\right) &= \bm{R}\bm{M}_{{\rm b},ij}^t\left(t\right)-k_{2}\Delta \bm{\omega}_{ij}^t\Delta t,
\end{align}
with $\bm{R}$ a rotation operator that aligns the previous bond force and moment with the contact plane at the current time, $\Delta \dot{\bm{x}}_{{\rm c},ij}$ the translational velocity difference between the particles at the contact point, and $\Delta \omega_{ij}$ the angular velocity difference between the particles (with the normal component being the rotation aligned with the contact plane, and the tangential component indicating the rotation perpendicular to the plane). $k_{1}$ denotes the nondimensional spring constant for the bond force, and $k_{2}$ indicates the corresponding constant for the bond moment. For computational efficiency we choose a first-order in time method in accordance with the original method outlined in \citet[][]{POTYONDY20041329}, and employ a sufficiently small time step to ensure convergence. The advantage of the current bond model lies in the fact that, for sufficiently stiff springs, we approximate a rigid aggregate that is "glued together" and settles as a single body. Further, by modeling a complex body as many elementary spherical shapes, the Lagrangian marker distribution is simplified compared to performing an IBM implementation on a single, irregularly-shaped body. In the current implementation we do not consider the breakage of bonds. This allows us to collapse the spring coefficients of the original model, which are based on the material properties and strength of the bonding material, to a single coefficient $k = k_{1} = k_{2}$, under the assumption that for rigidity to be ensured, we only need to make the springs sufficiently stiff, as will be discussed in further detail below.

\subsection{Initial and boundary conditions}
The simulations are initiated with the fluid at rest, and employ triply periodic boundary conditions. The aggregate is released from rest, with a horizontal orientation $\theta\left(t=0\right) = \theta_{0} = \pi/2$. When other initial orientations are used, this will be mentioned explicitly. The computational domain has size $W_{\rm c}\times H_{\rm c} \times W_{\rm c} = 10 D_{\rm L} \times 50 D_{\rm L} \times 10 D_{\rm L}$, which ensures that the influence of the boundaries remains minimal throughout the simulations. For some cases with large Galileo numbers, we increased the domain height up to $120 D_{\rm L}$, in order to allow the aggregate to reach a steady state without interacting with its own wake via the periodic boundaries in the vertical direction. Additionally, we chose a tall domain to be able to capture the entire wake generated by the aggregate. Due to the periodic boundary conditions in the vertical direction, the downward force of the particle acting on the fluid is not balanced and would lead to a constant downward acceleration of the fluid. To counteract this acceleration, we apply a corrective, distributed, upward force per volume $\bm{f}_{\rm adj}$ to the fluid that is uniformly distributed and equal and opposite to the negative buoyancy of the particles \citep{PhysRevE.61.7146}, which in its dimensionless form is 
\begin{equation}
    \bm{f}_{\rm adj} = -\frac{D_{\rm L}^3}{W_{\rm c}^2H_{\rm c}}\frac{\pi}{6}\left(1+\alpha^{-1/3}\right) \ .
\end{equation}

\subsection{Resolution and validation}
\label{sec:grid_res}

Validation and convergence information for the case of individual settling spheres can be found in the supplementary material, as well as in \citet{Bieg_dis}. For the current scenario of settling aggregates, we obtained some general guidance for the appropriate time step size $\Delta t$ by keeping the Courant–Friedrichs–Lewy (CFL) number
\begin{equation}
{\rm CFL} = \frac{u \, \Delta t}{h} \ 
\end{equation}
below 0.5, where $u$ indicates the maximum vertical velocity value within the domain at any given time. A sufficiently low CFL number ensures the time step to be small enough to prevent instability in the numerical solution, and to ensure the discretization converges towards an accurate solution of the case being studied. Beyond that, we further decreased the time step based on convergence studies, in order to ensure that any restrictions imposed by the bond model would not affect the numerical results, such that ${\rm CFL} = 0.1$. This resulted in typical time steps of order $\Delta t \sim 10^{-3}$.

When simulating a pair of bonded particles, the spatial and temporal resolutions require additional consideration. The spatial step size $h$ needs to be small enough to resolve the fluid-particle interactions, and the temporal step size $\Delta t$ has to be sufficiently small to capture the spring-like motion of the bond, whose dimensional period,
\begin{equation}
    T = 2\pi \sqrt{\frac{m_{\rm S}}{k}},
\end{equation}
is estimated based on the dimensional spring coefficient $k$ and the mass $m_{\rm S}$ of the smaller particle.

We follow the approach of \citet{vowinckel_withers_luzzatto-fegiz_meiburg_2019} by taking the grid spacing $h$ to be at most one-twentieth of the average particle diameter, ${(D_{\rm L}+D_{\rm S})}/{2} \geq 20\,h$. To fully resolve the spring-like behavior, we limit the time step to be at most 1/32 of the spring period. Tests showed the values to be sufficient to ensure both stability and convergence.

\begin{figure}
    \centering
        \subfigure[]{\includegraphics[width=0.49\textwidth]{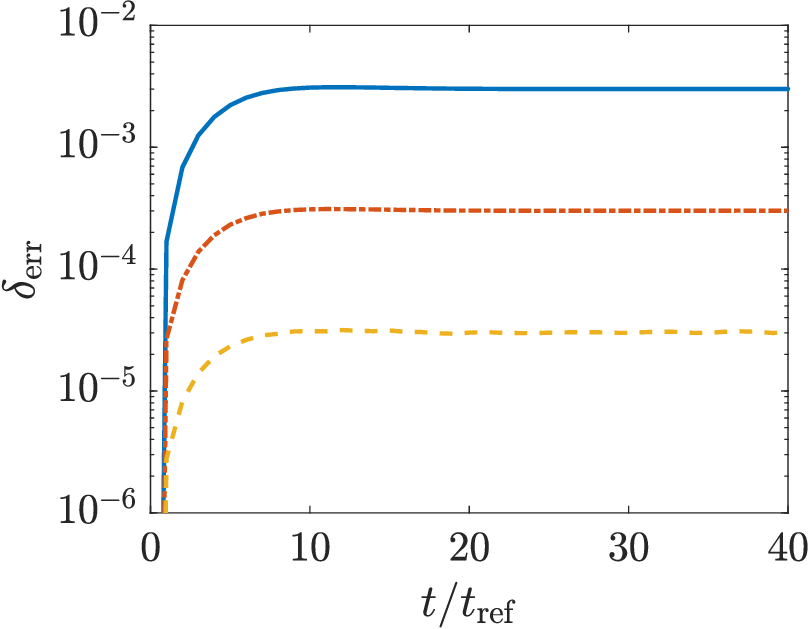}}
    \subfigure[]{\includegraphics[width=0.49\textwidth]{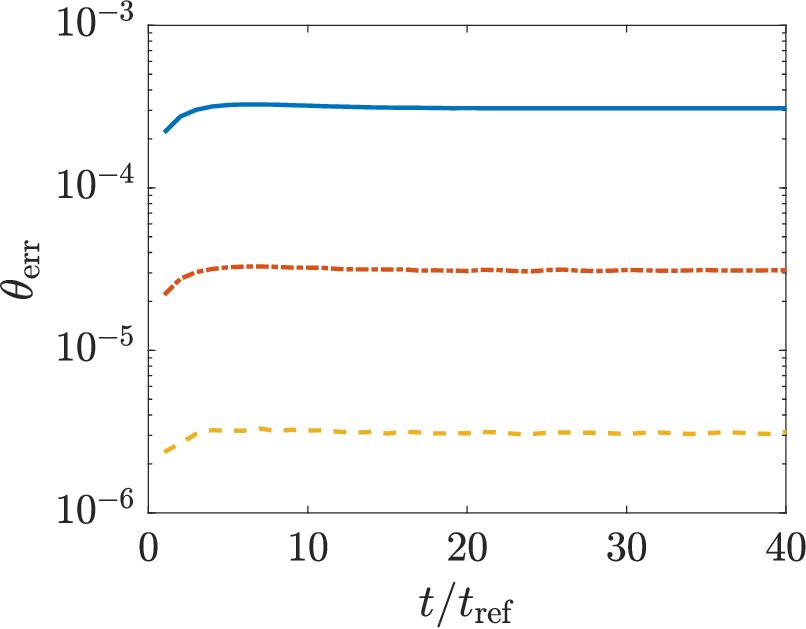}}
    \subfigure[]{\includegraphics[width=0.49\textwidth]{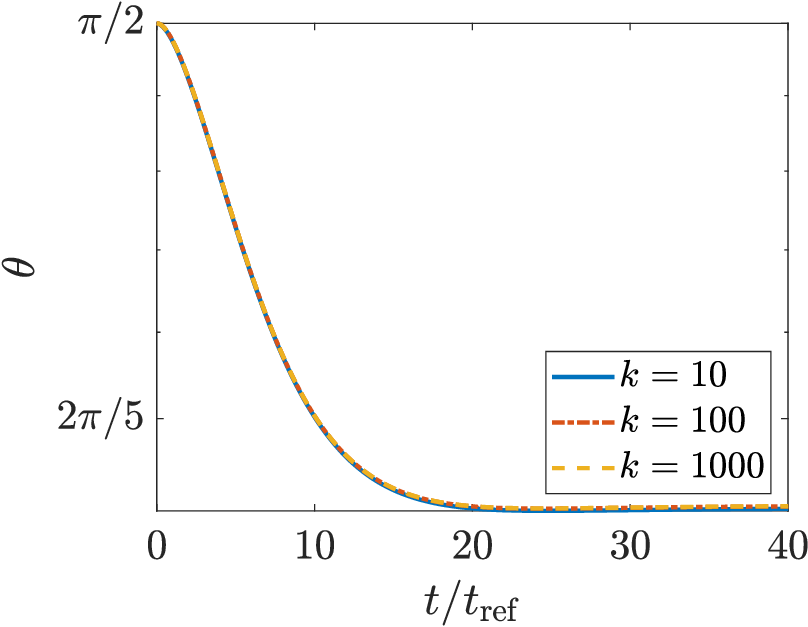}}
    \caption{Convergence of simulation results for increasing bond strength: measure of the error (a) $\delta_{\rm err}$ associated with the gap size between the particles, and (b) $\theta_{\rm err}$ associated with the bending of the aggregate. These error measures are shown as functions of time for $Ga = 13$ and $\alpha = 1.25$, and for several values $k$. (c) Corresponding orientation angles $\theta$.}
    \label{fig:b_strength}
\end{figure}

To determine a suitable computational bond strength coefficient $k$, we assess the deformation of the aggregate by tracking $\delta_{\rm err} = \left|\left|\bm{x}_{\rm L} - \bm{x}_{\rm S}\right|-\left(D_{\rm L}+D_{\rm S}\right)/2\right|$, for particle centers $\bm{x}_{\rm L}$ and $\bm{x}_{\rm S}$, and $\theta_{\rm err} = \left|\theta_{\rm L}-\theta_{\rm S}\right|$, for particle orientations $\theta_{\rm L}$ and $\theta_{\rm S}$, as functions of time. Figures \ref{fig:b_strength}(a) and \ref{fig:b_strength}(b) show that a value $k=1000$ suffices to keep these deformation measures below $10^{-4}$, which we determined to be sufficiently small compared to the characteristic length scale that it could be safely neglected. Figure \ref{fig:b_strength}(c) shows that the evolution of the aggregate's orientation is fully converged for this value of $k$, so that we select it for our simulations.

To validate the implementation of the bond between two spheres, we compare with the recent results of \citet[][]{nissanka_ma_burton_2023}, who considered the settling behavior of two connected spheres in a fluid of density $\rho_{\rm f} = 971\,{\rm kg \, m^{-3}}$ and viscosity $\nu = 0.01\,{\rm m^2 \, s^{-1}}$. The spheres have equal diameters $D = 0.002\, {\rm m}$, but different densities $\rho_1 = 1420\, {\rm kg \, m^{-3}}$ and $\rho_2 = 2790\, {\rm kg \, m^{-3}}$. In the following, we use the average particle density $\left(\rho_1+\rho_2\right)/2 = 2105 \, {\rm kg \, m^{-3}}$ for determining a representative Galileo number $Ga = 0.03$ based on equation (\ref{eq:re_term}), to be consistent with the notation of \citet[][]{nissanka_ma_burton_2023}. In the experiments, the two spheres were glued together and released from rest in a tank of dimensions $0.004 \, {\rm m}\times 0.19 \, {\rm m}\times 0.15 \, {\rm m}$. Initially, the denser particle was located approximately above the less dense one, so that the pair will rotate upon release until the denser particle is below the less dense one. In the experiment, the location of the contact point between the spheres, as well as the orientation of the pair, were recorded as functions of time.

For the simulation, we take a computational domain of $2D\times 40D \times 10D$, which matches the experiment in the $x$-direction and is sufficiently large in the $y$- and $z$-directions so that the boundaries do not influence the settling behavior. The boundaries themselves are modeled as no-slip walls. The experiments in \citet[][]{nissanka_ma_burton_2023} were performed with $Ga \sim O\left(10^{-2}\right)$, which is too small for our code to simulate, as it would require too small a time step. We instead perform several simulations for somewhat larger values of $Ga$, to show that we converge to the experimental results of \citet[][]{nissanka_ma_burton_2023} as $Ga$ approaches that of their experiments. We initialize the aggregate at an orientation angle of $17\pi/18$, so that the lighter particle is slightly displaced from being directly under the denser one. 

\begin{figure}
    \centering
        \subfigure[]{\includegraphics[width=0.49\textwidth]{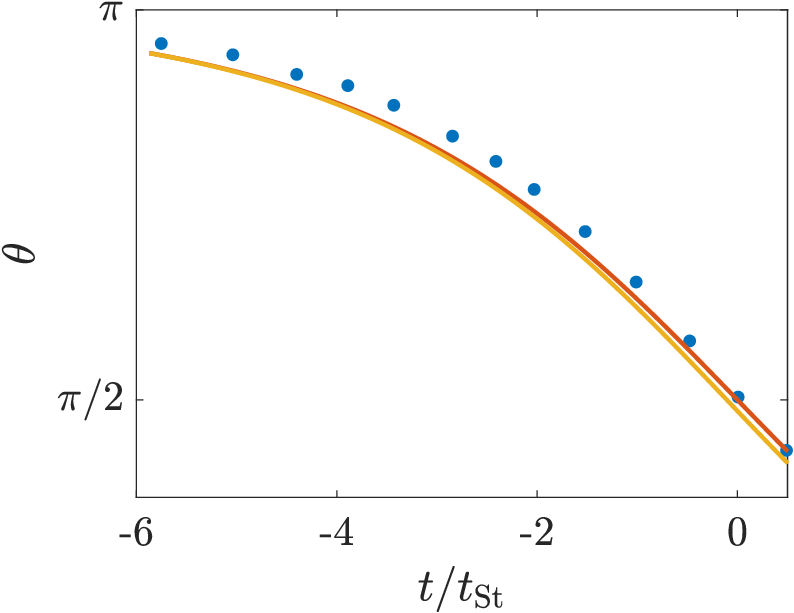}}
    \subfigure[]{\includegraphics[width=0.49\textwidth]{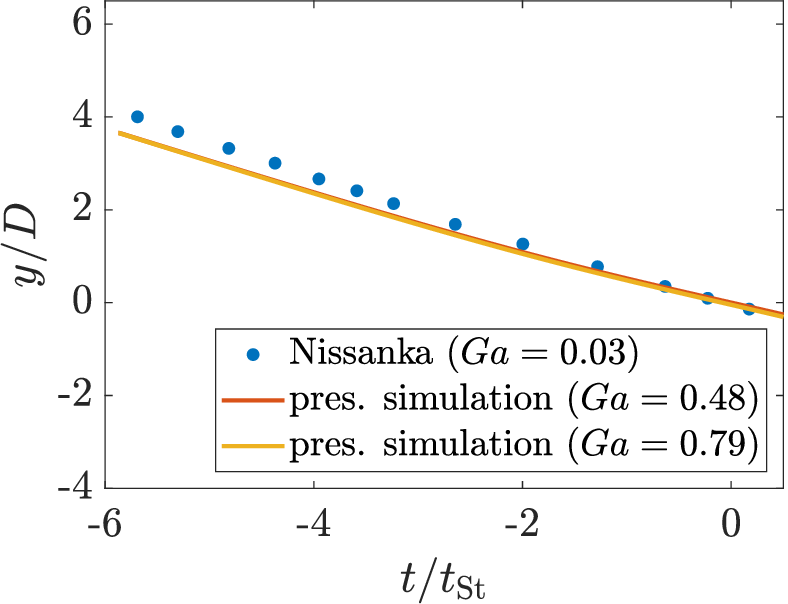}}
    \caption{Comparison with the experimental results of \citet[][]{nissanka_ma_burton_2023} (figure 3 in their article), for the settling of an aggregate consisting of two spheres of equal size and different densities: (a) orientation angle $\theta$ and (b) vertical position $y/D$. All results are non-dimensionalized by using the particle diameter $D$ as the length scale and $t_{\rm St} = D/u_{\rm St}$ as the time scale, where the Stokes settling velocity $u_{\rm St}$ is calculated using the density of the lighter particle.}
    \label{fig:burton2}
\end{figure}

Figure \ref{fig:burton2} shows the orientation angle $\theta$ of the aggregate, along with the vertical position $y/D$ of the geometrical center of the pair ($t=0$ is chosen to be where $\theta = \pi/2$) for the simulations, as well as the corresponding experimental data of \citet[][]{nissanka_ma_burton_2023}. Here time is scaled by the reference time $t_{\rm St} = D/u_{\rm St}$, where $u_{\rm St}$ denotes the Stokes settling velocity of the lighter sphere. We find that even when $Ga$ is slightly higher than the value of the experiments, when the system is scaled in reference to the Stokes settling velocity the results collapse onto the experimental results, which further validates the computational bond model used in the following.

%%%%%%%%%%%%%%%%%%%%%%%%%%%%%%%%%%%%%%%%%%%%%%%%%%%%%%%%%%%%%
%%%%%%%%%%%%%%%%%%%  Results and discussion %%%%%%%%%%%%%%%%
%%%%%%%%%%%%%%%%%%%%%%%%%%%%%%%%%%%%%%%%%%%%%%%%%%%%%%%%%%%%

\section{Results and discussion}\label{sec:num_only}

We begin by exploring the parameter range $Ga \in [8,28]$ and $\alpha \in [1,2]$, for which the experiments most clearly demonstrate the existence of a steady-state terminal settling velocity and orientation. Following comparisons between experiments and simulation results in this regime, we will expand the discussion to the broader range of $Ga = O(1)-O(10^2)$ and $\alpha \leq 4$.

\subsection{Temporal evolution}
\begin{figure}
    \centering
    \subfigure[]{\includegraphics[width=0.49\textwidth]{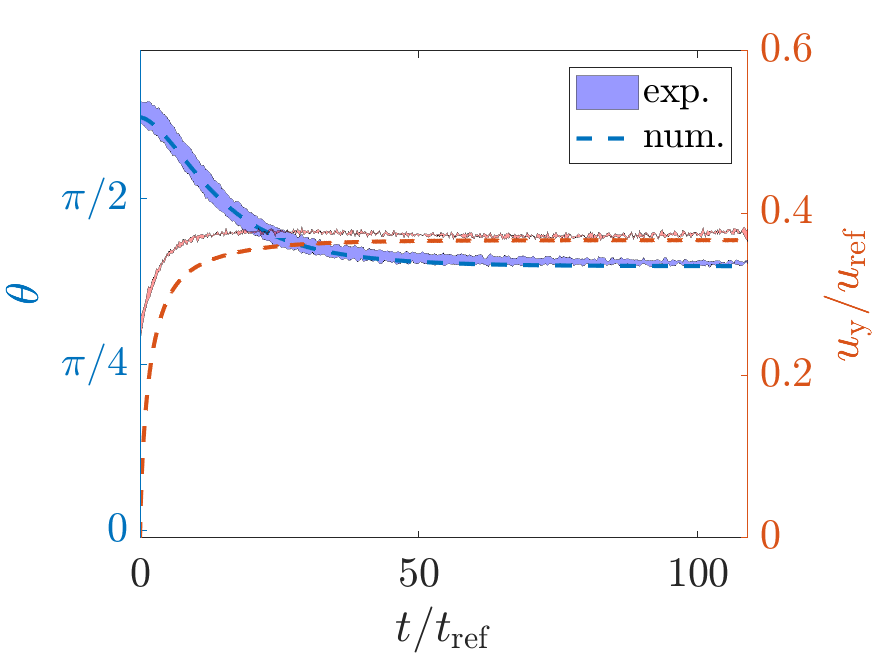}}
        \subfigure[]{\includegraphics[width=0.49\textwidth]{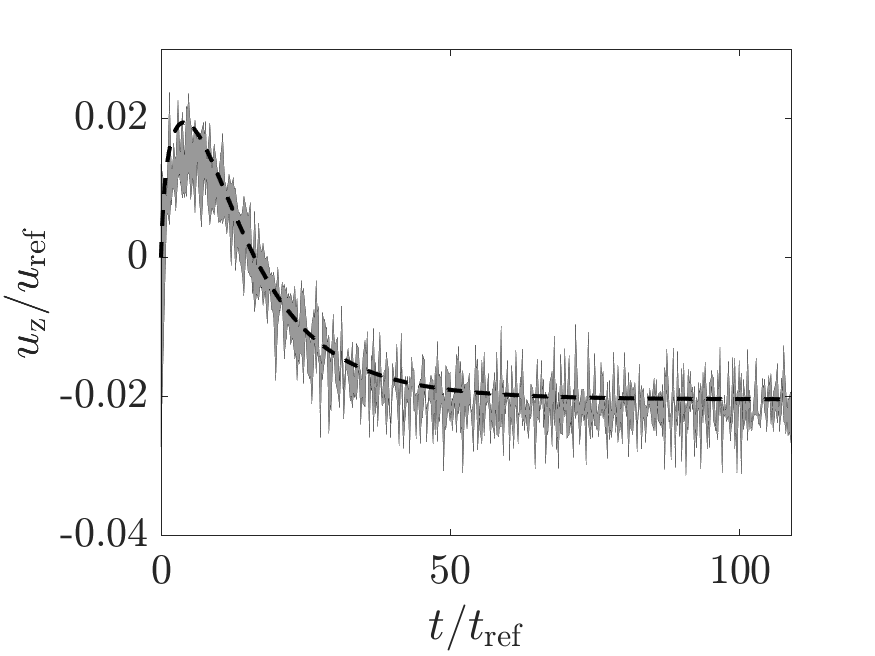}}
    \caption{Comparison of numerical and experimental data for $\alpha = 8/7$, $Ga = 15$. (a) reports the orientation angle $\theta$ (blue) and the vertical velocity $u_y/u_{\rm ref}$ (red). (b) shows the horizontal velocity $u_z/u_{\rm ref}$ (black). Dashed lines indicate the numerical results, while the shaded region indicates the experimental uncertainties of one standard deviation around the average experimental result.}
    \label{fig:exp_sing}
\end{figure}

Figure \ref{fig:exp_sing} compares the temporal evolution of the experimental and numerical settling behavior. In the simulations, the aggregates are released from rest, with an orientation angle $\theta_{0}$ equal to the initial value observed in the corresponding experiments (note that for experiments, there is likely always a small initial rotation and velocity generated during the release of the aggregate. The vertical settling velocity $u_{y, {\rm term}}/u_{\rm ref}$ and the horizontal drift velocity $u_{ z, {\rm term}}/u_{\rm ref}$ of the center of mass of the pair of particles are recorded as functions of time, along with the orientation angle $\theta$. The figure demonstrates good agreement between simulation results and experimental observations, both during the initial transient stage and for the subsequent steady state. Figure \ref{fig:exp_sing}(b) shows that the horizontal drift velocity generally points in the direction of the lower, larger sphere. We remark that for a given pair of $\alpha$ and $Ga$ values, the initial orientation of the aggregate in the simulation has no influence on the final, steady-state settling velocity or orientation.

\begin{figure}
    \centering
    \subfigure[]{\includegraphics[width=0.49\textwidth]{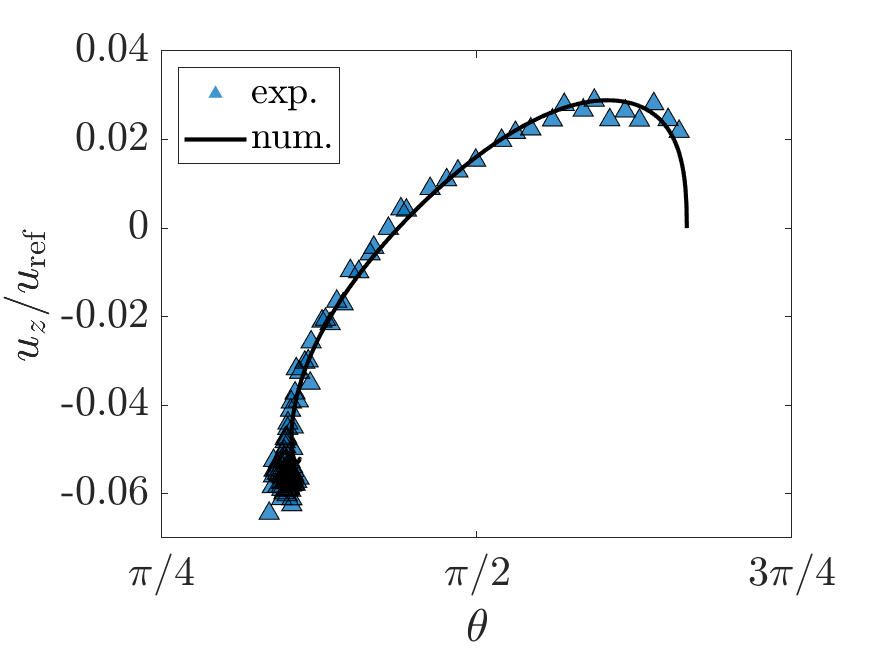}}
        \subfigure[]{\includegraphics[width=0.49\textwidth]{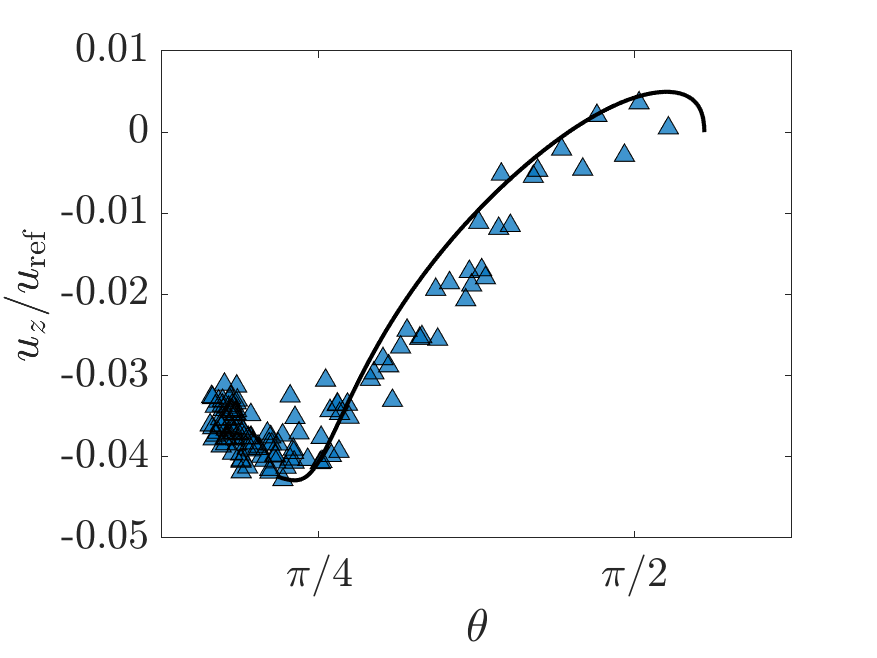}}
     \caption{Instantaneous rescaled drift velocity $u_{\rm z}/u_{\rm ref}$ as a function of the orientation angle $\theta$, showing both experimental data (symbols) and simulation results (line). Each plot depicts many time steps of a single experiment and simulation, for (a) $Ga = 21$, $\alpha = 1.4$, $\theta_{0} = 2\pi/3$ and (b) $Ga = 23$, $\alpha = 2$, $\theta_{0} = 5\pi/9$. The largest drift velocities are generally observed for orientations $\theta$ near $\pi/4$, while zero horizontal velocity corresponds to the symmetric orientation $\theta = \pi/2$.}
    \label{fig:exp_uzth}
\end{figure}

Figures \ref{fig:exp_uzth}(a)-(b) show comparisons between experimental and simulation results for the instantaneous drift velocity as a function of the orientation angle for different sets of parameters $Ga$, $\alpha$, and $\theta_{0}$. The figures demonstrate good agreement between numerical and experimental data, and highlight the fact that the largest horizontal drift velocities are observed for orientation angles near $\theta=\pi/4$.

\begin{figure}
    \centering
    \subfigure[]{\includegraphics[width=0.49\textwidth]{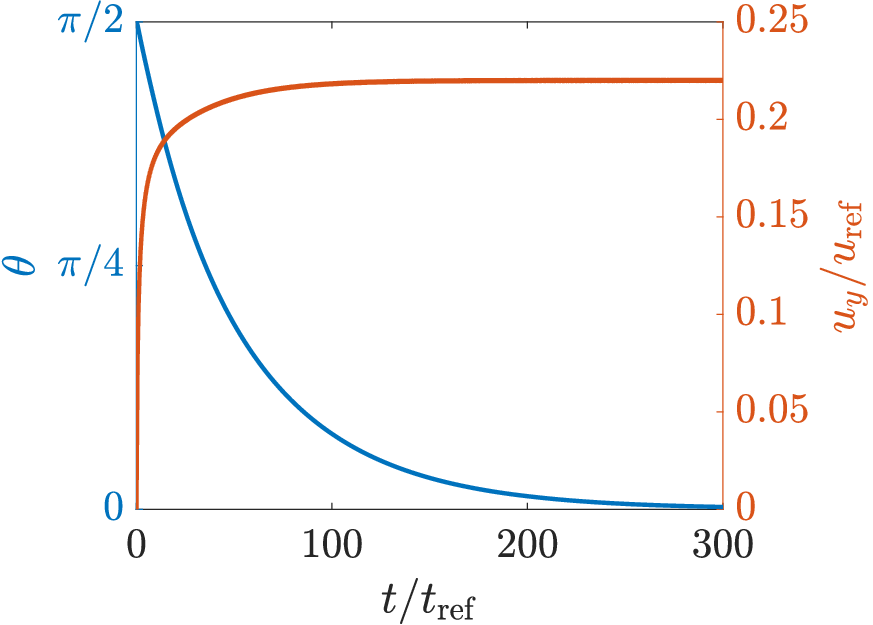}}
    \subfigure[]{\includegraphics[width=0.49\textwidth]{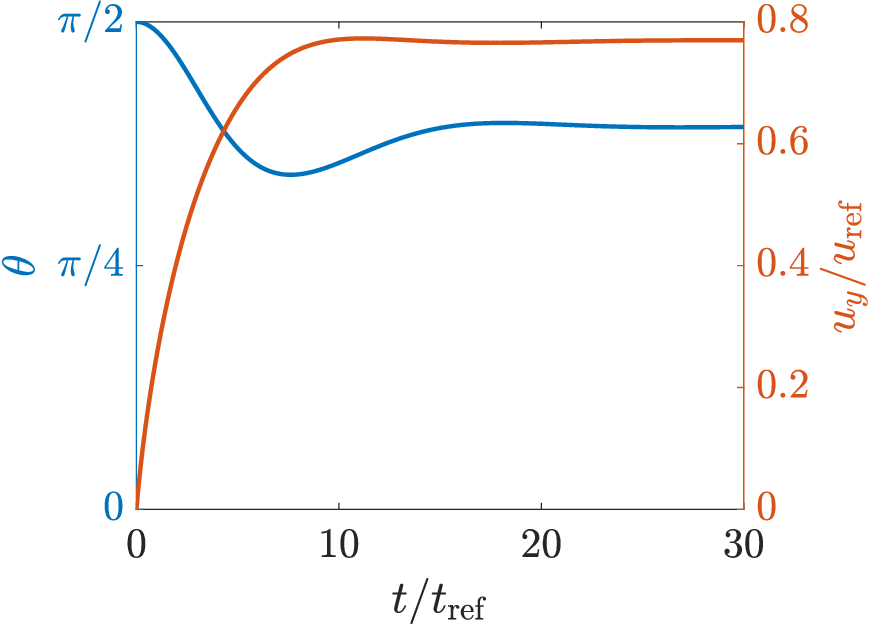}}
    \caption{Numerical simulation results for the time evolution of the orientation angles $\theta$ and the velocity component $u_y/u_{\rm ref}$ as functions of time for $\alpha = 1.5$ and (a) $Ga=4$, and (b) $Ga = 42$. For the larger $Ga$-value the aggregate behaves as an underdamped oscillator.}
    \label{fig:uz_th}
\end{figure}

Figure \ref{fig:uz_th} (a)-(b) compares simulation results for two different values of $Ga$ but identical $\alpha = 1.5$. For the lower $Ga$-value, the orientation angle $\theta$ decreases monotonically with time until the aggregate becomes aligned in the vertical direction, (figure \ref{fig:uz_th}(a)). For the larger value of $Ga$, reported in figure \ref{fig:uz_th}(b), the orientation angle no longer evolves monotonically and the terminal orientation is nonvertical. Here the aggregate behaves like an underdamped oscillator, in that it overshoots its equilibrium orientation and settles into a steady state through damped oscillations due inertia. Again, we observe that the steady state is reached more quickly for the larger Galileo number. These results suggest that for larger values of $Ga$ the aggregate tends to approach its terminal state in an oscillatory fashion, and the terminal state is less likely to be vertically oriented. We will return to these points in more detail in the following.

\begin{figure}
    \centering
\subfigure[]{\includegraphics[width=0.5\textwidth]{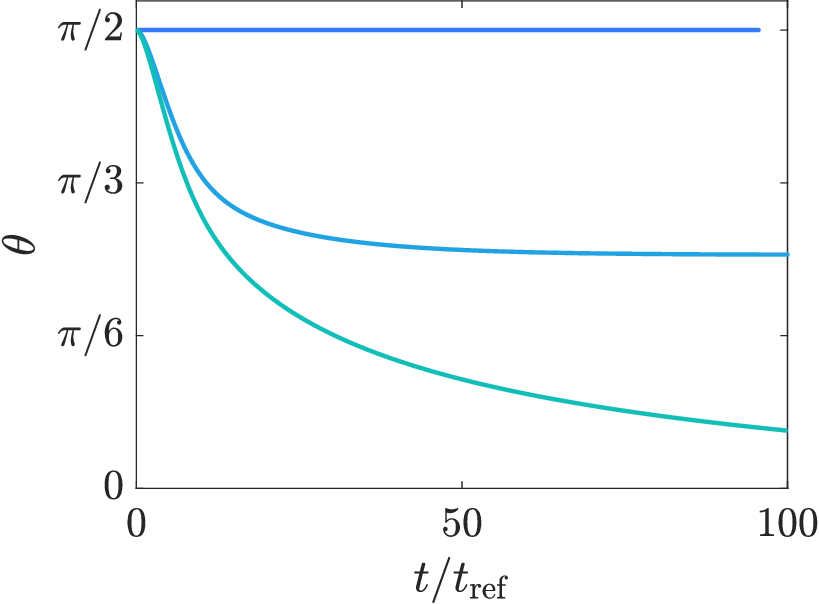}}\subfigure[]{\includegraphics[width=0.48\textwidth]{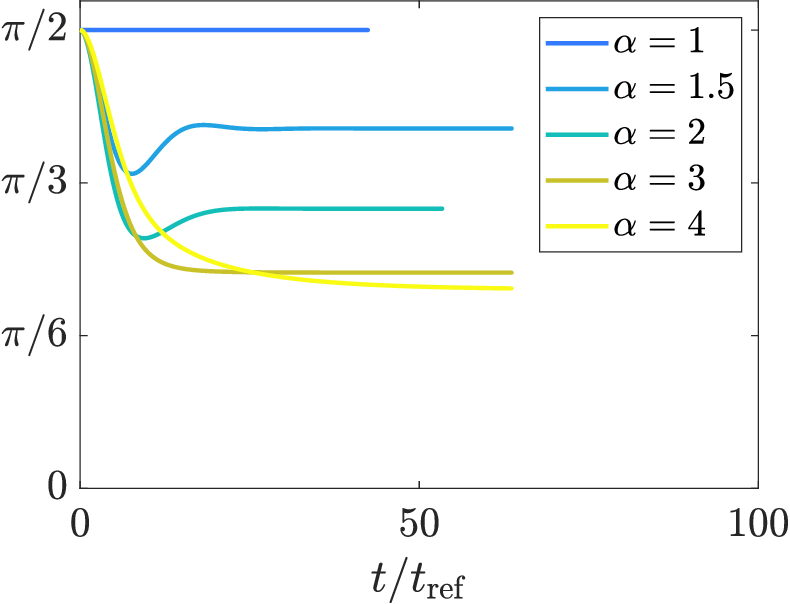}}
    \caption{Orientation angle of the aggregate over time, for varying values of $\alpha$ and fixed Galileo numbers: (a) $Ga = 13$, and (b) $Ga = 42$. The time required to converge to a steady state depends on both $\alpha$ and $Ga$.}
    \label{fig:timescale}
\end{figure}

Figures \ref{fig:timescale}(a) and \ref{fig:timescale}(b) show the temporal evolution of the orientation $\theta$ of the aggregate as $\alpha$ is varied, for two different values of the Galileo number. Generally, larger values of $\alpha$ result in terminal orientations closer to the vertical direction, although at smaller $Ga$ a vertical orientation is achieved for much smaller values of $\alpha$ than at larger $Ga$. At lower Galileo numbers, the aggregates again take significantly longer to rotate into their terminal configuration than at higher Galileo numbers. Indeed, in the example case at $Ga=13$ shown in figure \ref{fig:timescale}(a), the length of the transient phase increases with $\alpha$, especially when the final orientation is close to vertical ($\theta = 0$).

\begin{figure}
    \centering
\subfigure[]{\includegraphics[width=0.5\textwidth]{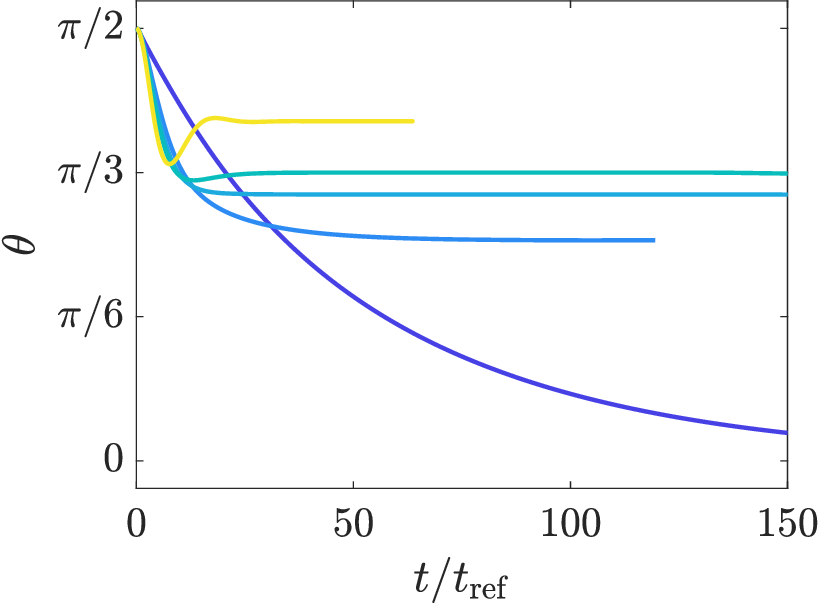}}\subfigure[]{\includegraphics[width=0.48\textwidth]{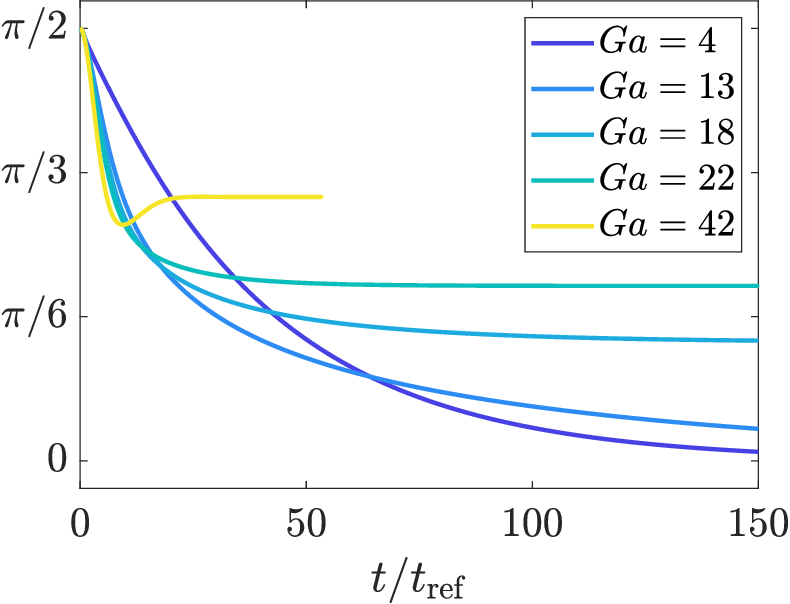}}
    \caption{Orientation angle of the aggregate as a function of time, for various Galileo numbers and (a) $\alpha = 1.5$ and (b) $\alpha = 2$. For a constant value of $Ga$, a larger $\alpha$ leads to a more vertical terminal orientation.}
    \label{fig:timescale2}
\end{figure}

The evolution of $\theta$ with time as $Ga$ is varied is shown in figures \ref{fig:timescale2}(a) and \ref{fig:timescale2}(b), for two different values of the diameter ratio $\alpha$. The time to reach the terminal state usually decreases for larger $Ga$, in agreement with the previous paragraph. We note that this observation does not hold for certain cases where the terminal orientation approaches the vertical, such as $\alpha=2$ with $Ga = 13$. For the same $\alpha$, a smaller $Ga$ generally leads to a more vertically aligned terminal state.

\begin{figure}
    \centering
    \subfigure[]{\includegraphics[width=0.49\textwidth]{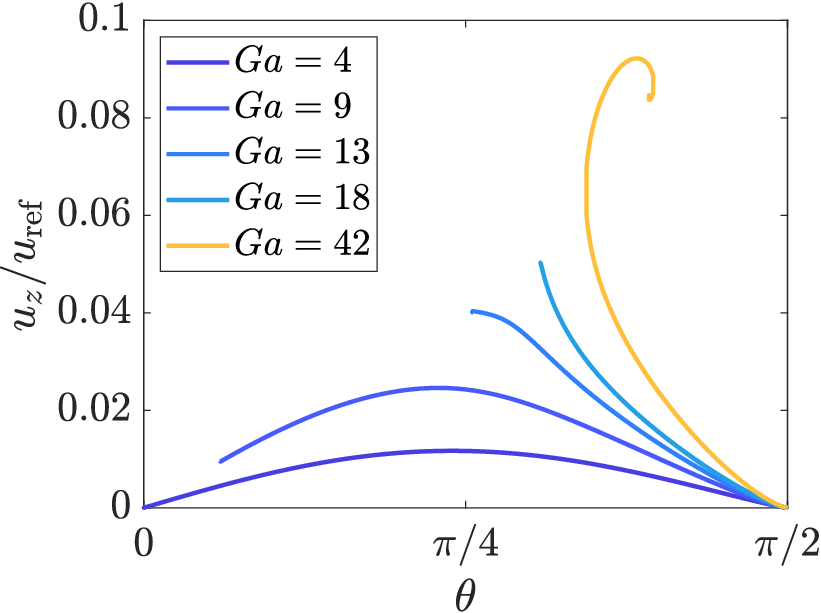}}\subfigure[]{\includegraphics[width=0.49\textwidth]{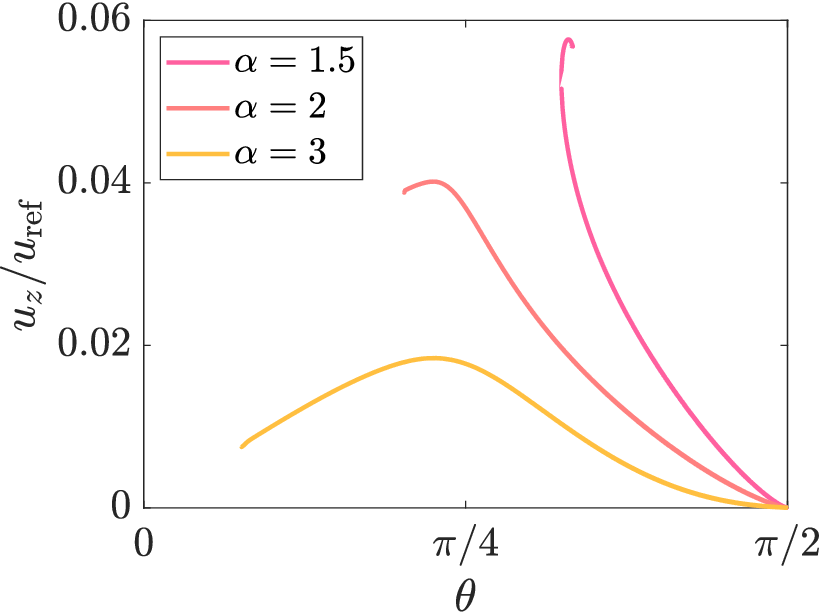}}
    \caption{Horizontal drift velocity as a function of $\theta$, for (a) different values of $Ga$ and $\alpha = 1.5$, and (b) $Ga = 22$ and various values of $\alpha$. Varying $Ga$ and $\alpha$ lead to significant variation in the magnitude of the maximum drift velocity, as well as the value of $\theta$ corresponding to that maximum.}
    \label{fig:theta_uz}
\end{figure}

Figures \ref{fig:theta_uz}(a) and \ref{fig:theta_uz}(b) show the time-dependent horizontal drift velocity $u_z/u_{\rm ref}$ as a function of the orientation angle $\theta$, for different values of $Ga$ and $\alpha$. For a constant $\alpha$, we find that the maximum drift velocity increases with $Ga$, while for constant $Ga$, it increases with decreasing $\alpha$. The emerging spiral shapes seen for the largest values of $Ga$ and the smallest values of $\alpha$ reflect the underdamped oscillations described above.

\begin{figure}
    \centering
    \subfigure[]{\includegraphics[width=0.49\textwidth]{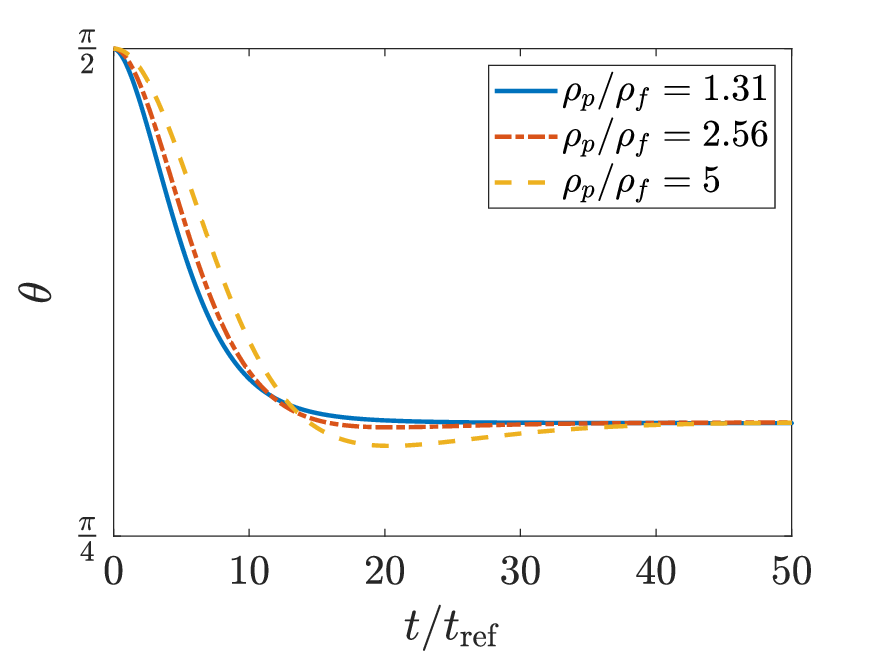}}\subfigure[]{\includegraphics[width=0.49\textwidth]{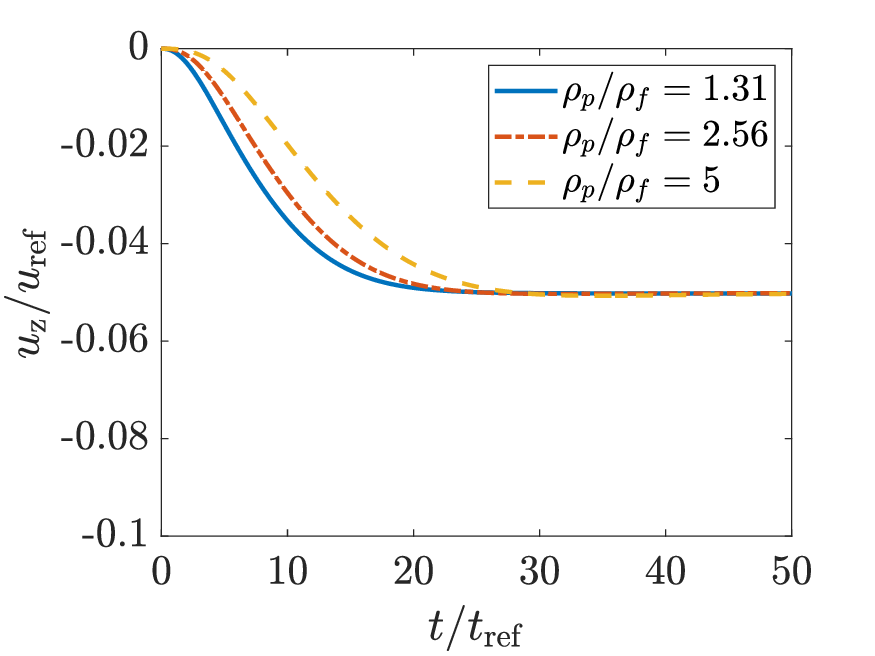}}
    \caption{Orientation angle (a) and horizontal settling velocity (b) for aggregates with different density ratios $\rho'$, for $Ga = 18$ and $\alpha = 1.5$. While $\rho'$ modifies the transient dynamics, it does not affect the terminal settling properties.}
    \label{fig:density}
\end{figure}

We also vary the density ratio $\rho'$, to elucidate its effect on the settling behavior. In figure \ref{fig:density} we show simulations for aggregates with different $\rho'$, while keeping $Ga$ fixed. We find that altering $\rho'$ only influences the transient behavior, while the terminal settling properties do not depend on $\rho'$.

In summary, all of the above combinations of $\alpha$ and $Ga$ demonstrate the emergence of a terminal state characterized by steady values of the settling and drift velocities, and of the orientation angle. Lower Galileo numbers $Ga$ and larger aspect ratios $\alpha$ favor a monotonic evolution of these quantities towards their terminal values, whereas larger $Ga$ and smaller values of $\alpha$ are seen to promote the emergence of underdamped oscillations. Having considered the transient evolution of the aformentioned values, we consider in the following section the terminal behavior.

\subsection{Terminal behavior}

\begin{figure}
    \centering
\includegraphics[width=0.495\textwidth]{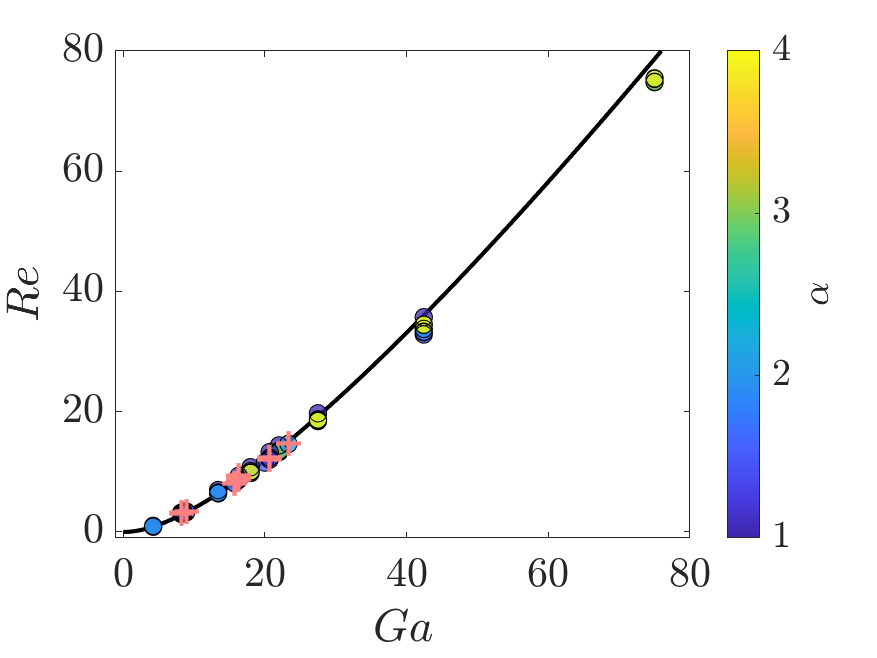}
    \caption{Relationship between $Ga$ and $\Rey$: The solid line represents equation (\ref{eqn:re_ar}) derived by \citet{NGUYEN199753}. Filled circles show simulation results. The colorbar indicates the value of $\alpha$, and + symbols represent experimental data.}
    \label{fig:res_reref}
\end{figure}

To compare some aspects of our experimental and simulation results with the existing literature, it is useful to discuss our data in terms of both the Galileo number $Ga$, which is formed with the buoyancy velocity (cf. equation (\ref{eq:re_term})), and the Reynolds number $\Rey$, which is based on the terminal settling velocity (cf. equation (\ref{eq:re})). Based on a series of experiments for a single spherical particle, \citet{NGUYEN199753} proposed an empirical relationship between $Re$ and $Ga$:

\begin{equation}\label{eqn:re_ar}
    \Rey = \left[\frac{18}{Ga^2}+\frac{3}{16}\left(1+0.079Ga^{1.498}\right)^{-0.755}\right]^{-1}.
\end{equation}

Figure \ref{fig:res_reref} demonstrates that, across the parameter range that we investigated, this empirical relationship captures qualitatively well our experimental and numerical results, including for different values of $\alpha$. We further notice that there seems to be a nearly single-valued relationship between $Ga$ and $\Rey$, despite $Ga$ not being affected by changes in $D_{\rm S}$ (and by extension the overall aggregate mass) while $\Rey$ does vary with $D_S$, since the presence of the smaller sphere modifies the settling velocity. This reflects the fact that in the present range $Ga \in \left[4,75\right]$, the settling velocity is a weak function of $\alpha$. Hence, for this range of $Ga$ we can use equation (\ref{eqn:re_ar}) to compare the results as functions of $Re$ and $Ga$, respectively.

\begin{figure}
    \centering
\subfigure[]{\includegraphics[width=0.49\textwidth]{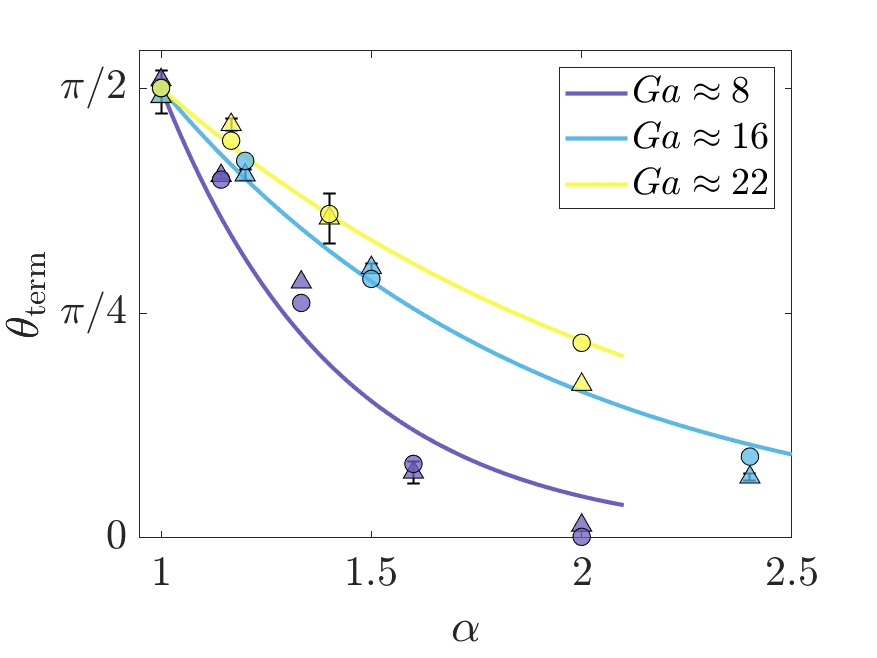}}\subfigure[]{\includegraphics[width=0.49\textwidth]{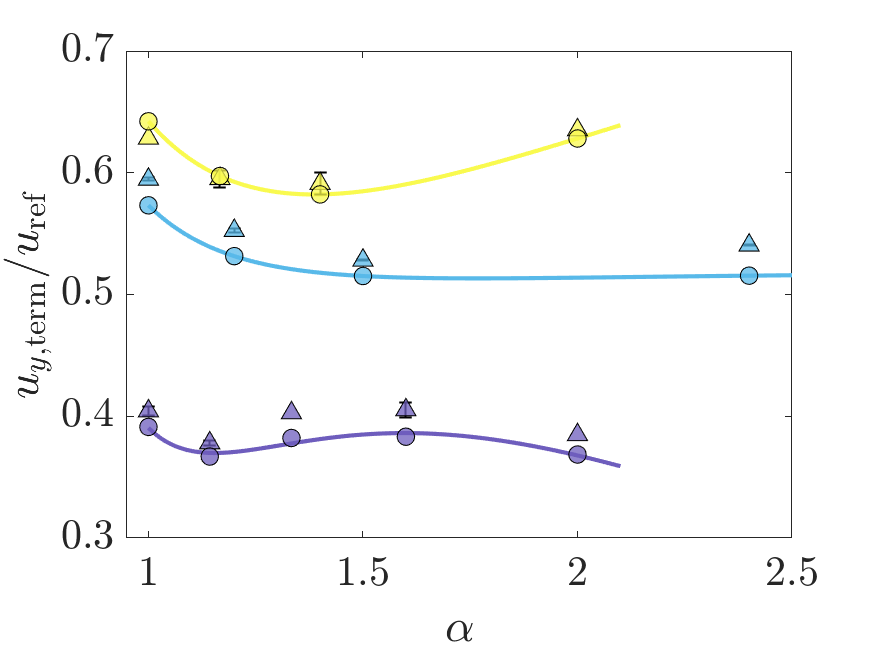}}
\subfigure[]{\includegraphics[width=0.49\textwidth]{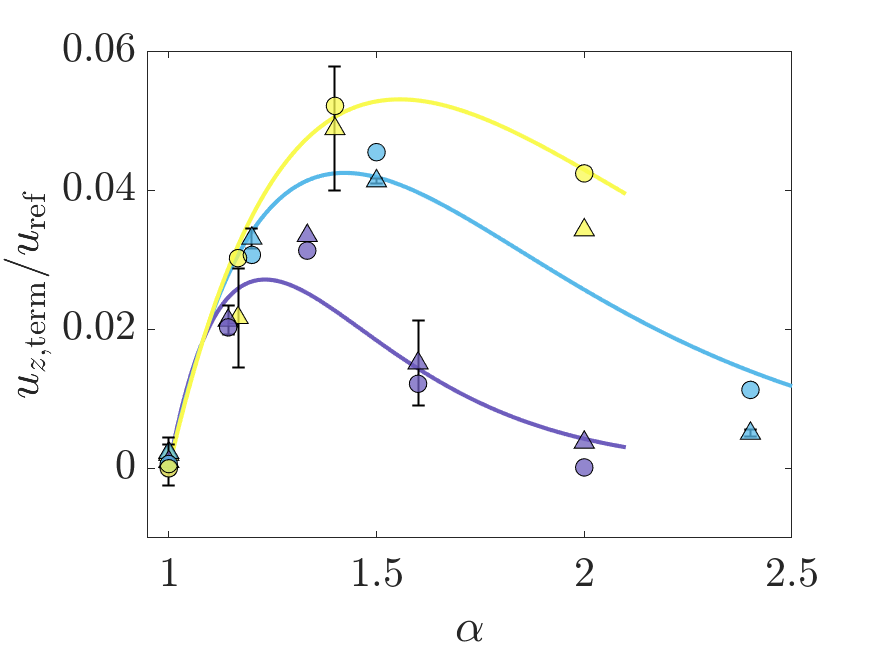}}
    \caption{Comparison between experimental ($\bigtriangleup$ symbols with error bars) and numerical (filled circles) results for (a) the terminal orientation angle, (b) the settling velocity, and (c) the drift velocity, for various values of $Ga$ and $\alpha$. Colors indicate average $Ga$ values for experiments with similar $Ga$, as the Galileo number could not be fixed to an exact value between experiments for different $\alpha$ values. Details regarding the best-fit curves are provided in the text.}
    \label{fig:key_exp}
\end{figure}

We now focus on the terminal values of the orientation angle, $\theta_{\rm term}$, as well as the corresponding settling and drift velocity components, $u_{y, {\rm term}}$ and $u_{ z, {\rm term}}$. In figure \ref{fig:key_exp}, we compare the terminal orientation angle and the settling and drift velocities measured experimentally to their numerical counterparts, for varying values of $Ga$ and $\alpha$. 

Within the parameter range explored here, we observe that all three terminal variables vary strongly with $Ga$ and $\alpha$. In figure \ref{fig:key_exp}(a), both simulations and experiments demonstrate a pronounced decrease of $\theta_{\rm term}$ from $\theta_{\rm term} = \pi/2$ when increasing $\alpha$ at approximately (taking into account variations in the experimental viscosity as discussed in section \ref{sec:exp}) constant $Ga$, indicating that more unequal sizes of sphere tend to align the aggregates increasingly in the vertical direction (i.e. $\theta_{\rm term} = 0$). We also note that the horizontal settling orientation seen for $\alpha = 1$ corresponds to the results of \citet[][]{khayat_cox_1989}, which show that objects with fore-aft symmetry will align themselves horizontally with the flow. Figure \ref{fig:key_exp}(b) suggests that for $Ga \approx 22$ the terminal settling velocity of the aggregates decreases with $\alpha$ for small values of $\alpha$, reaches a minimum for an intermediate value $\alpha \approx 1.4$, and subsequently grows again. For $Ga \approx 8$ and 16, the variation of the settling velocity with $\alpha$ evolves similarly but is less pronounced than for higher $Ga$. Interestingly, figure \ref{fig:key_exp}(c) demonstrates that the terminal horizontal drift velocity increases with $\alpha$ for small $\alpha$, then reaches a maximum, and subsequently decreases as $\alpha$ grows further. This is consistent with the assumption that as $\alpha \rightarrow \infty$ the effect of the smaller particle on the settling velocity of the aggregate becomes negligible so that the aggregate behaves similarly to a single settling sphere, which has no horizontal terminal velocity component.

The solid lines in figure \ref{fig:key_exp} represent least squares fits of the simulation results for the terminal variables as functions of $\alpha$ for a fixed value of $Ga$, which were obtained as follows. In the experiments, $Ga$ cannot be fixed to an exact value as $\nu$ varies due to room temperature variations. Hence, no two simulations plotted in figure \ref{fig:key_exp} have the exact same $Ga$: the fittings were instead performed for an average $Ga$ calculated from nearby values of $Ga$. For the terminal orientation angle, we assume that the relationship between $\theta_{\rm term}$, $\alpha$, and $Ga$ can be fitted by the form
\begin{align}\label{eq:theta_eqn}
    \theta_{\rm term}\left(\alpha,Ga\right) &= \left[\pi/2-\theta_{\rm \infty} \left(Ga\right)\right]e^{-A \left(Ga\right)\left(\alpha-1\right)}+\theta_{\rm \infty} \left(Ga\right),
\end{align}
where $\theta_{\rm \infty}(Ga)$ indicates the orientation angle as $\alpha \rightarrow \infty$ as a function of $Ga$, and $A \left(Ga\right)$ is a fitting parameter different for each value of $Ga$. For the terminal settling velocity, $u_{y, {\rm term}}/u_{\rm ref}$, we expect that it will converge to the settling velocity of a single sphere of diameter $D_{\rm L}$, $u_{\rm L, term}$, as $\alpha \rightarrow \infty$, so that we fit a polynomial of the form
\begin{align}
    \frac{u_{y, {\rm term}}\left(\alpha,Ga\right)}{u_{\rm ref}} &= \frac{u_{\rm L, term}}{u_{\rm ref}}+C_1\left(Ga\right)\alpha^{-3}+C_2\left(Ga\right)\alpha^{-2}+C_3\left(Ga\right)\alpha^{-1} ,\label{eq:uyref}
\end{align}
to the data. Here $C_i\left(Ga\right)$ denote fitting parameters.

As mentioned above, we expect the terminal drift velocity $u_{ z, {\rm term}}$ to vanish for $\alpha = 1$ and in the limit $\alpha \rightarrow \infty$. Hence we assume a relationship of the form
\begin{equation}
    \frac{u_{z, {\rm term}}\left(\alpha,Ga\right)}{u_{\rm ref}} = J_1(Ga) \left(\alpha-1\right)e^{-J_2(Ga)\left(\alpha-1\right)} , \label{eq:uzref}
\end{equation}
where $J_i\left(Ga\right)$ are fitting parameters. While other correlations may be possible, we chose this relation based on our observations of the settling aggregate for the values of $Ga$ and $\alpha$ considered.

\begin{figure}
    \centering
    \subfigure[]{\includegraphics[width=0.49\textwidth]{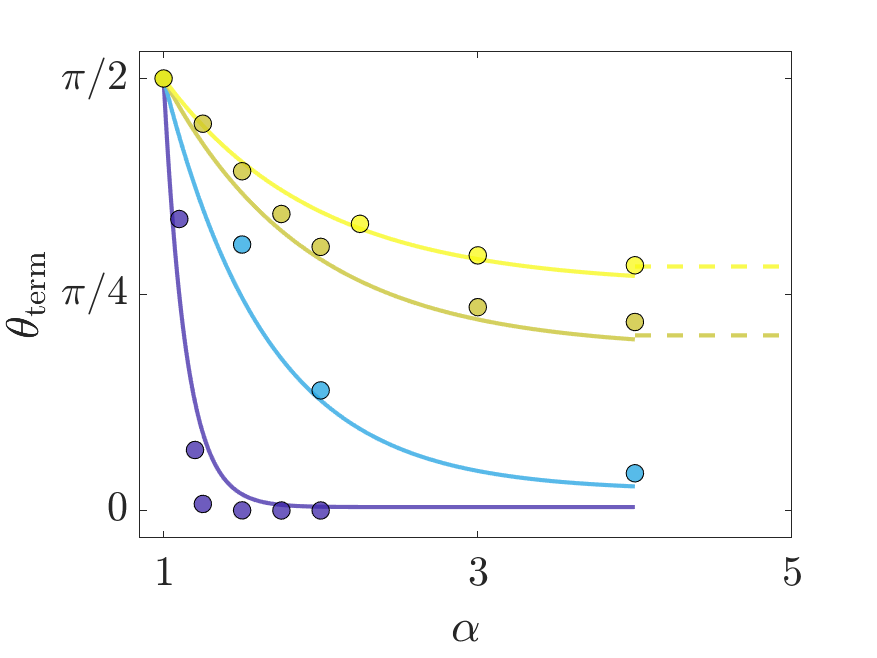}}
    \subfigure[]{\includegraphics[width=0.49\textwidth]{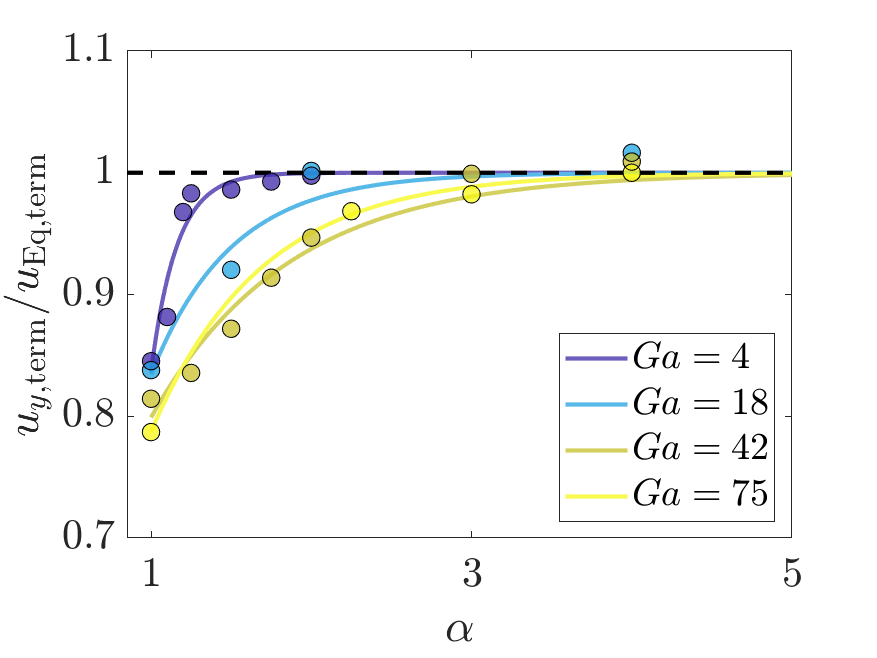}}
    \subfigure[]{\includegraphics[width=0.49\textwidth]{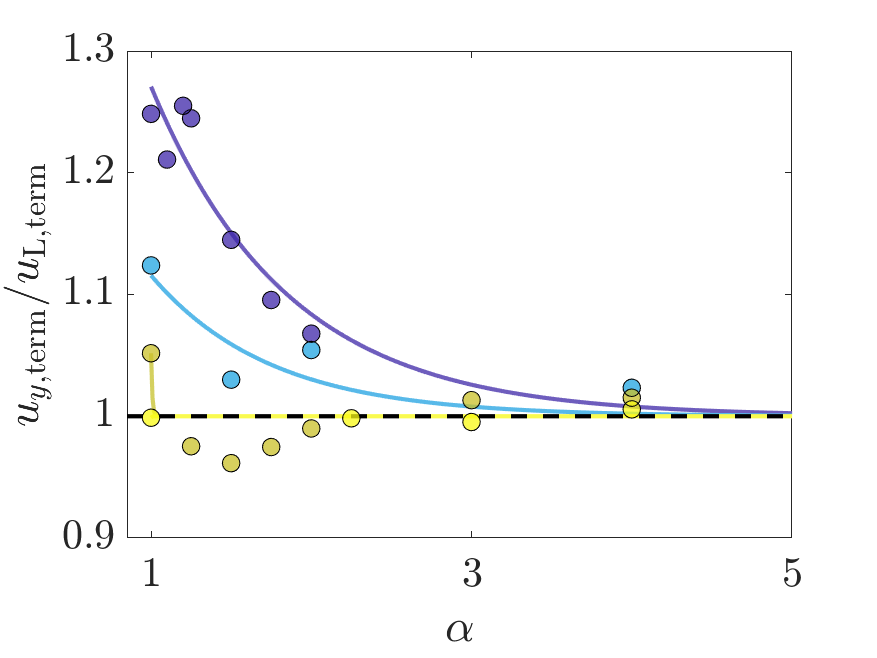}}
    \subfigure[]{\includegraphics[width=0.49\textwidth]{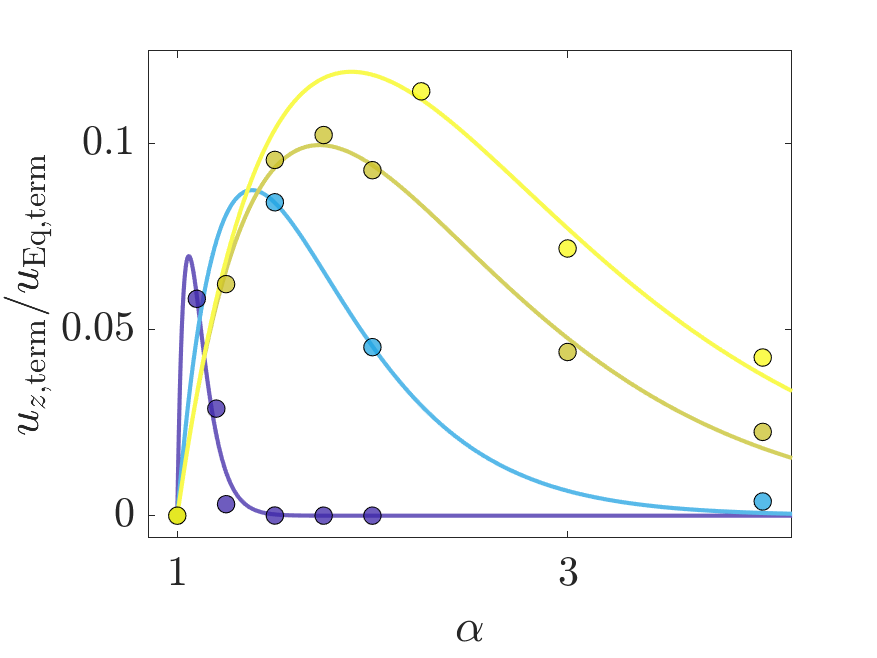}}
    \caption{Numerical simulation results for the (a) terminal orientation angle, (b) vertical velocity normalized by the velocity of a sphere of the same volume as the aggregate, (c) vertical velocity normalized by the velocity of the larger sphere alone, and (d) horizontal drift velocity normalized by the settling velocity of a sphere with the same volume as the aggregate, for various values of $Ga$ and $\alpha$. The solid lines represent empirical fits, as explained in the main text.}
    \label{fig:a_values}
\end{figure}

The above comparisons show that the results of the numerical simulation agree well with the experimental data for the parameter regime covered by the experiments. This provides the necessary validation to explore a wider parameter range through the numerical simulations alone, as shown in figure \ref{fig:a_values}(a)-(d). The results of the numerical simulations shown in figure \ref{fig:a_values}(a) indicate that for $Ga\lesssim 25$, the terminal orientation angle asymptotically approaches the vertical direction as $\alpha \rightarrow \infty$. As $Ga$ increases within this range, this asymptotic convergence with $\alpha$ slows down, and outside of this range, the terminal orientation displays the emergence of a finite-valued plateau for $\alpha \approx 4$.
The dynamics within this range of relatively small $Ga$ are qualitatively similar to those derived theoretically by \citet[][]{candelier_mehlig_2016} for two spheres connected by a massless, infinitely thin rod settling at small, but finite $Re$. If the rod length is much larger than the particle diameters, the authors find
\begin{equation}
\theta_{\rm term}\left(\alpha,\Rey\right) = \left\{
        \begin{array}{ll}
           \frac{\pi}{2}-\pi \, {\rm arcsin}\left(\frac{16\alpha\left(\alpha-1\right)}{3\Rey}\right) + O\left(\left(\frac{\alpha^2}{D_{\rm L}}-\frac{\alpha}{D_{\rm L}}\right)\delta x\right)& \quad \alpha < \alpha_{\rm 0} \\
            0 & \quad \alpha \geq \alpha_{\rm 0}
        \end{array},
    \right.
    \label{eq:mehlig}
\end{equation}
where
\begin{equation}
    \alpha_{\rm 0} = \frac{1+\sqrt{1+\frac{3}{4}\Rey}}{2} \; ,
    \label{eq:alphac}
\end{equation}
where $\delta x$ is the dimensionless distance between particle centers. The equation is valid when $\delta x \Rey/\alpha \ll 1$.

While we are able to qualitatively capture the decrease in the orientation angle using equation (\ref{eq:mehlig}), the prediction overshoots the numerical results for the present Galileo numbers. This overshoot can be attributed to the difference in $\Rey$ values from the region where equation (\ref{eq:mehlig}) holds, and the absence in our simulations and experiments of the long connecting rod assumed in deriving this equation.

We now return to the observation that for $Ga \gtrsim 25$, figure \ref{fig:a_values}(a) indicated that the orientation angle of the aggregate no longer approaches the vertical direction $\theta_{\rm term} = 0$ for $\alpha \gg 1$. Instead, the aggregate takes an inclined orientation that is tilted further away from the vertical as $Ga$ increases. We propose the following explanation for these non-zero terminal orientation angles: as $Ga$ (or $Re$) increases, the flow begins to separate near $Ga \approx 28$ ($\Rey \approx 20$) for our geometry. \citet[][]{bub_drop_part} observe that for a smooth sphere, the separation angle varies with $Re$ as
\begin{equation}
\theta_{\rm sep}\left(\Rey\right) = 
        \begin{array}{ll}
            0.236\pi{\rm ln}\left(\frac{Re}{20}\right)^{0.438} \, , & \quad 20 < \Rey < 400.
        \end{array}\label{eq:sep_ang}
\end{equation}

The dashed horizontal lines in figure \ref{fig:a_values}(a) represent the values of $\theta_{\rm sep}$ for $Ga =$ 42 and 75, given by equations (\ref{eqn:re_ar}) and (\ref{eq:sep_ang}). For sufficiently large $Ga$, when the flow separates, and the aggregate no longer takes a vertical orientation for large $\alpha$, the aggregate instead aligns itself such that the smaller sphere is located near the separation line.

\begin{figure}
    \centering
            \begin{tabular}{cc}
\includegraphics[width=0.49\textwidth]{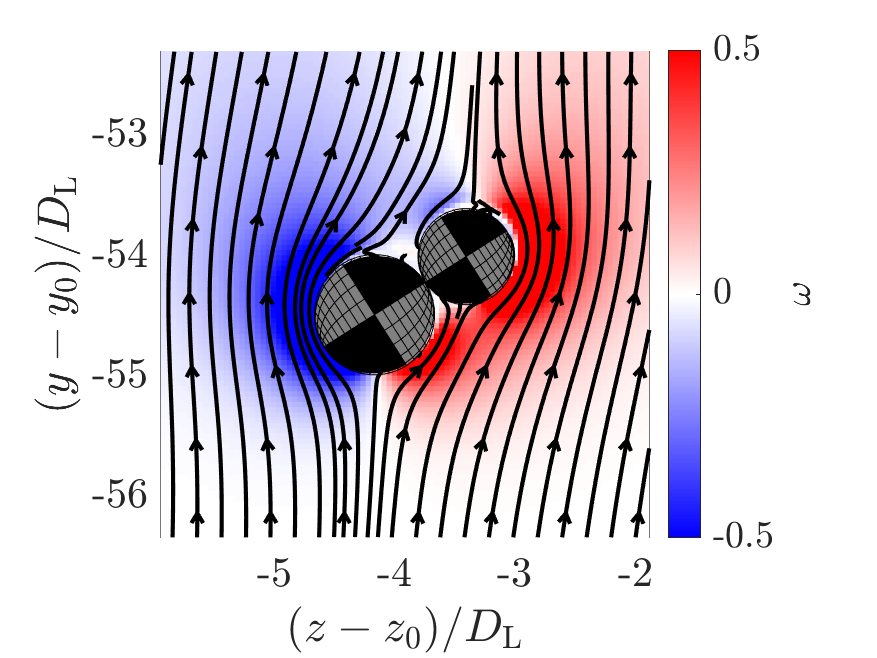} &   \includegraphics[width=0.49\textwidth]{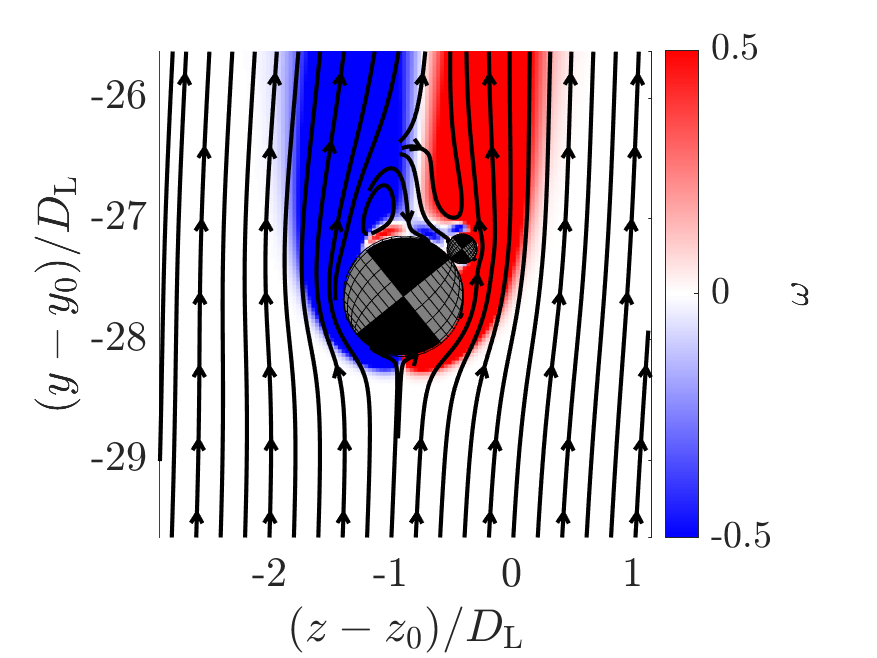} \\
\textbf{(a)} & \textbf{(b)} \\
\includegraphics[width=0.4\textwidth]{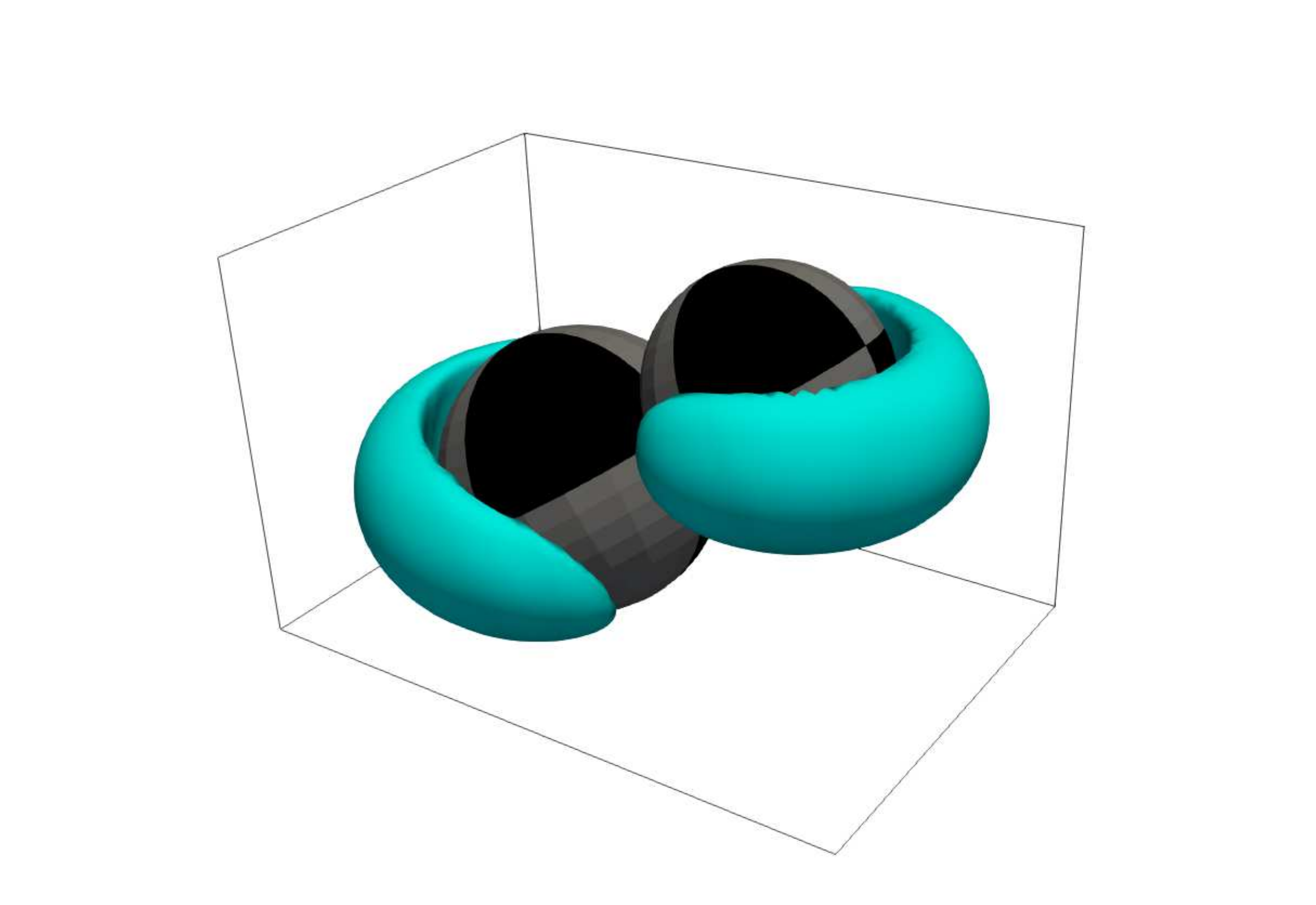} & \includegraphics[width=0.4\textwidth]{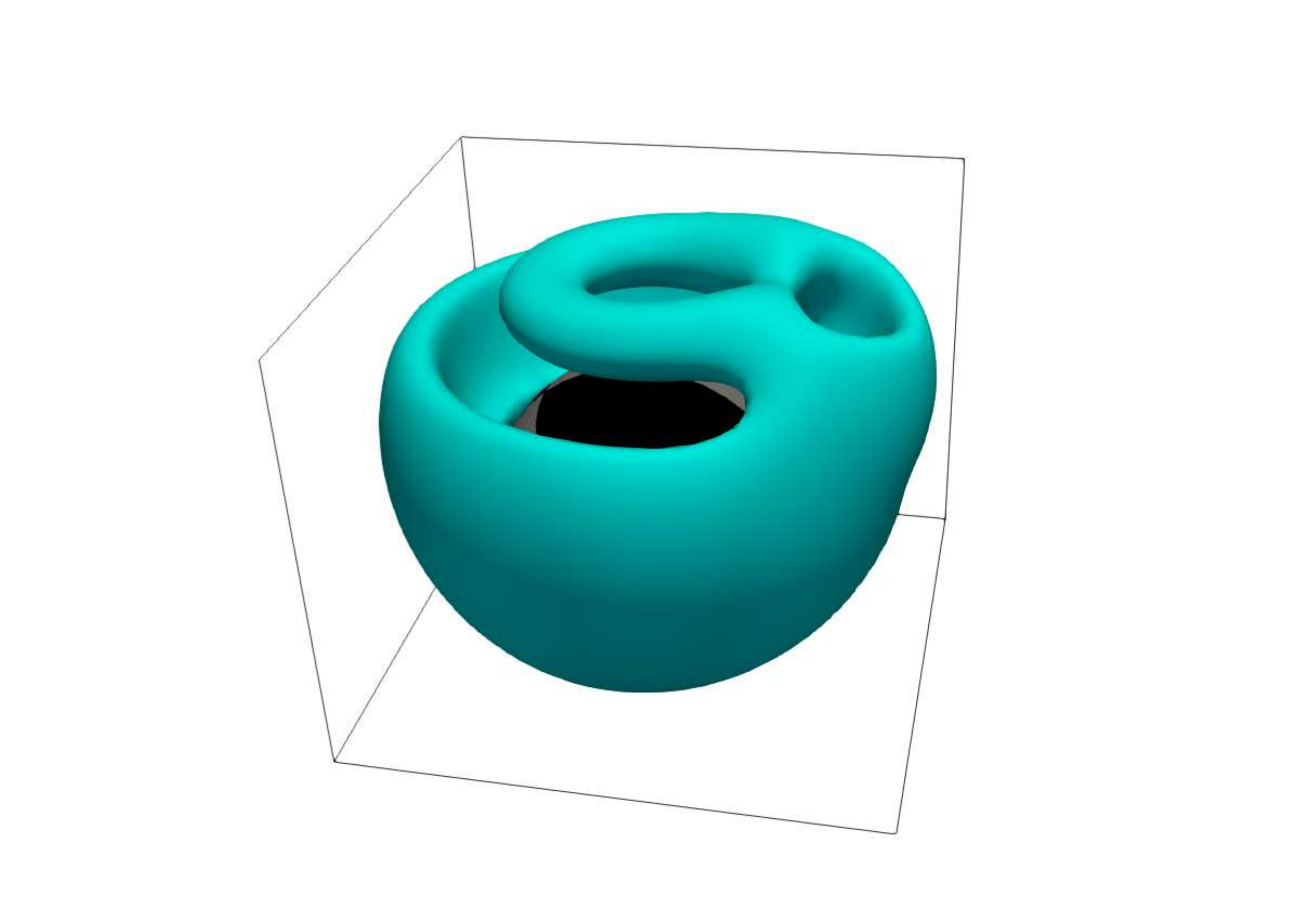} \\
\textbf{(c)} & \textbf{(d)} \\
\end{tabular}    
\caption{(a)-(b): Steady-state vorticity fields and streamline patterns in the reference frame moving with the aggregate, shown in the symmetry plane. (c)-(d): Contours of $Q$, with the value of $Q$ chosen to best demonstrate the characteristic vortical structures around the aggregate. The center of mass of the aggregate is initially positioned at $(y_0,z_0)$. Panels (a) and (c) are intermediate orientations for $Ga=9$, $\alpha = 1.25$, $Q=0.08$, and (b) and (d) for $Ga=75$, $\alpha = 4$, $Q = 0.1$. The colorbar indicates the vorticity.}
    \label{fig:vort}
\end{figure}

This is demonstrated in figures \ref{fig:vort}(a)-(b), which show the streamline pattern in the reference frame moving with the aggregate, along with the vorticity field, in the symmetry plane for various combinations of $Ga$ and $\alpha$. In addition, the geometry of the dominant vortical structures is visualized by means of the $Q$-criterion \citep[]{chakraborty2005relationships}, where the quantity
\begin{equation}
    Q = \frac{1}{2}\left(||{\bm \Omega}||^2-||{\bm S}||^2\right)
\end{equation}
is evaluated from the vorticity vector ${\bm \Omega}$ and the strain rate tensor ${\bm S}$. Regions with $Q>0$ are those where the magnitude of the vorticity is greater than that of the rate of strain in the flow.

The streamlines and vorticity fields reflect the symmetry of the aggregate, as shown in figure \ref{fig:vort}. As $Ga$ increases, the vortical region emanating from the larger sphere extends further into the wake, and for $\alpha \gg 1$ it completely engulfs the smaller sphere, as shown in figure \ref{fig:vort}(b). Interestingly, this figure also shows that for $Ga=75$ the flow separates from the rear of the larger sphere, with the smaller sphere positioning itself at the separation point, as discussed earlier.

  \ref{fig:a_values}(b)-(c) also compare the terminal settling velocity of the aggregate to those of a sphere with the same volume as the aggregate ($u_{\rm Eq, term}$, figure (b)) and of the larger sphere alone ($u_{\rm L, term}$, figure (c)), with the single sphere values obtained via corresponding simulations. For small to moderate $\alpha$, the aggregate settles more slowly than a single sphere of equal volume, but faster than the larger sphere alone. When $\alpha$ becomes large, both ratios of the velocity asymptotically approach unity, as expected. However, the rate at which the respective ratios converge to unity depends strongly on $Ga$. The numerical data for the settling velocity ratios shown in figures \ref{fig:a_values}(b) and \ref{fig:a_values}(c) are captured well by functions of the form
\begin{align}
    \frac{u_{y, {\rm term}}\left(\alpha,Ga\right)}{u_{\rm Eq, term}\left(Ga\right)} &= 1-C_1e^{-C_2\left(\alpha-1\right)}\label{eq:uyeq}
\end{align}
and
\begin{align}
    \frac{u_{y, {\rm term}}\left(\alpha,Ga\right)}{u_{\rm L, term}\left(Ga\right)} &= 1+C_3e^{-C_4\left(\alpha-1\right)} \, ,\label{eq:uyl}
\end{align}
respectively, where the $C_{i}\left(Ga\right)$ represent fitting parameters. 

Figure \ref{fig:a_values}(d) shows the terminal drift velocity, normalized by $u_{\rm Eq, term}$. We note that for $Ga \gtrsim 25$ the aggregate retains a significant terminal drift velocity even for values of $\alpha$ as large as $\alpha = 4$. This is consistent with our earlier observation that in this parameter range, the aggregate orients itself such that the smaller particle is located at the separation line, and therefore the terminal orientation remains asymmetric. As $\alpha \rightarrow \infty$, however, we expect the terminal drift velocity to decay to zero (given an infinitesimally small sphere will have little effect on the overall aggregate), in spite of the asymmetric orientation, so that it can be captured by a function of the form
\begin{equation}
    \frac{u_{z, {\rm term}}\left(\alpha,Ga\right)}{u_{\rm Eq, term}\left(Ga\right)} = J_1\left(\alpha-1\right)e^{-J_2\left(\alpha-1\right)},\label{eq:uzhigh}
\end{equation}
where $J_1\left(Ga\right)$ and $J_2\left(Ga\right)$ are fitting parameters. We also note that equation (\ref{eq:uzhigh}) gives a maximal terminal drift velocity for $\alpha_{\rm drift,max} = 1/J_2\left(Ga\right)+1$ where it is equal to $u_{\rm drift,max}/u_{\rm y,Eq,term} = J_1/\left(J_2e\right)$. 

We furthermore note that, for aggregates with intermediate orientation angles $0 <\theta<\pi/2$, a region of larger pressure exists on the side of the larger particle directly below the smaller one. This generates a horizontal pressure gradient across the large sphere, which is the primary reason for the horizontal drift velocity.

\begin{figure}
    \centering
    \subfigure[]{\includegraphics[width=0.49\textwidth]{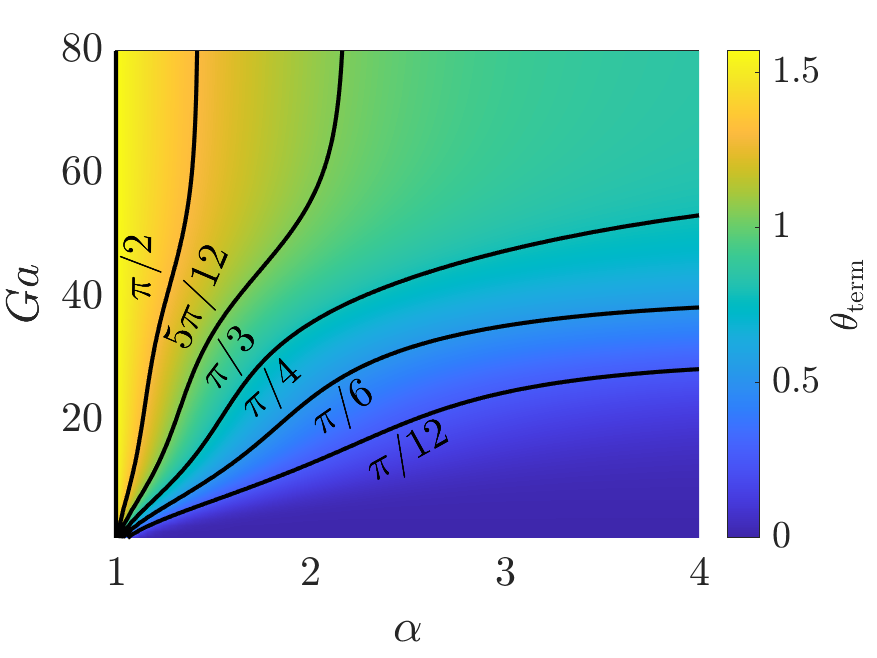}}\subfigure[]{\includegraphics[width=0.49\textwidth]{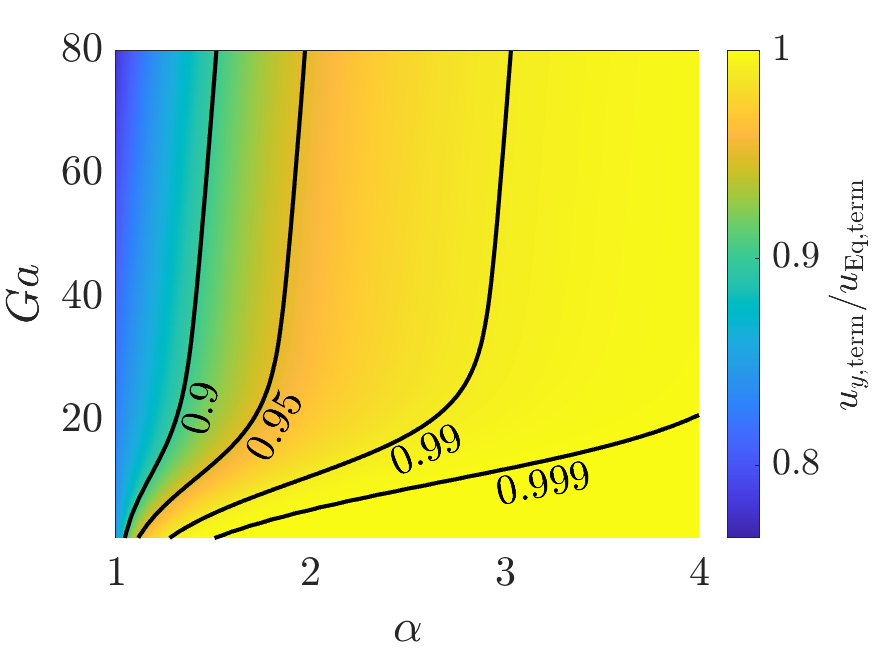}}
    \subfigure[]{\includegraphics[width=0.49\textwidth]{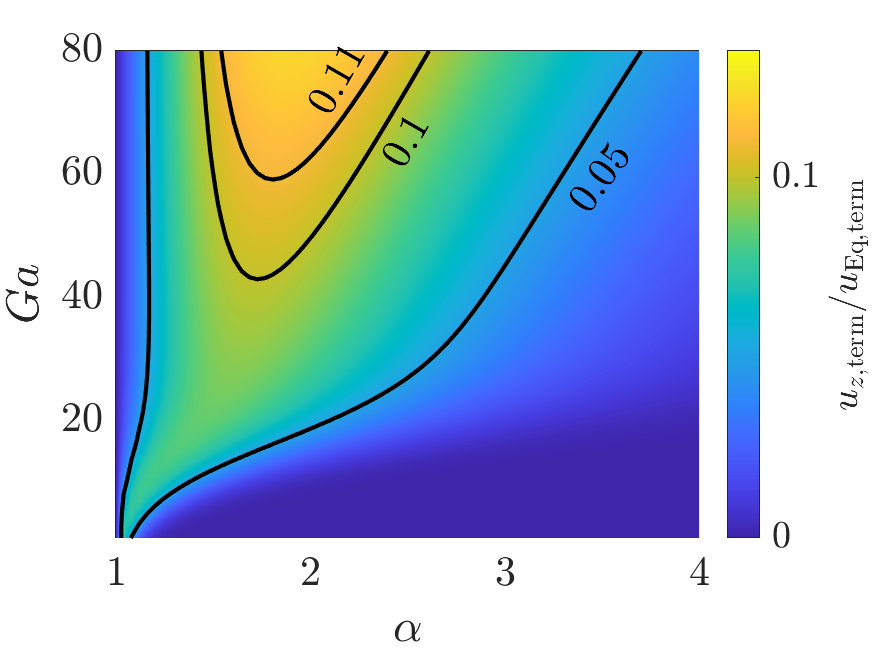}}
    \caption{Summary of the dependence of the (a) terminal orientation angle of the aggregate, (b) terminal settling velocity, and (c) terminal drift velocity on a ($Ga$,$\alpha$) phase diagram, based on the fits described in the text. The dashed line indicates the location of the maximum drift velocity. Angles are plotted in degrees for visual convenience.}
    \label{fig:ang_values}
\end{figure}

Figures \ref{fig:ang_values}(a)-(c) summarize the dependence of $\theta_{\rm term}$, $u_{\rm y,term}/u_{\rm Eq, term}$, and $u_{z, {\rm term}}/u_{\rm Eq, term}$ on $\alpha$ and $Ga$. To plot these figures, we fit $\theta_{\rm \infty}\left(Ga\right)$ in equation (\ref{eq:theta_eqn}) by means of a hyperbolic tangent, and $A\left(Ga\right)$ via a power law
\begin{align}
    \theta_{\rm \infty}\left(Ga\right) &= 0.412\left(1+\rm tanh\left(0.067\left(Ga-35.6\right)\right)\right),\\ 
    A\left(Ga\right) &= \frac{196}{\left(Ga+1.46\right)^2} +1.
\end{align}

These empirical fits match well with the experiments and the numerical simulations, deviating on average by approximately $0.11$ radians (7\% of a full $\pi/2$ rotation). Similarly, we obtain for the fitting parameters in equation (\ref{eq:uyeq}) 
\begin{align}
    C_1\left(Ga\right) &= 0.001Ga+0.155,\\
    C_2\left(Ga\right) &= 1.71e^{-0.189Ga+1.79}+1.52.
\end{align}
We find that the average deviation for the fitted model is $2\%$ of a single settling sphere's velocity $u_{\rm Eq, term}$. For the horizontal velocity $u_{z, {\rm term}}/u_{\rm Eq, term}$, we obtain fits for the parameters in equation (\ref{eq:uzhigh}) of the form
\begin{align}
    J_1\left(Ga\right) &= 0.233e^{-0.189Ga+3.31}+0.371,\\
    J_2\left(Ga\right) &= \frac{J_1\left(Ga\right)}{0.001Ga+0.074}e^{-1},
\end{align}
with the average deviation being $1\%$ of a single settling sphere's velocity $u_{\rm Eq, term}$. As such, these fits capture the behavior of the settling aggregates in the parameter space of $Ga$ and $\alpha$ considered, allowing us to predict the settling of small, two-particle aggregates settling in a fluid.

\section{Conclusion}\label{sec:conclusion}

We have investigated the settling dynamics of a model aggregate made of a pair of rigidly connected spherical particles of unequal size. By means of experiments and particle-resolved simulations, we have obtained a detailed picture of both the transient evolution and the terminal values of the orientation angle, settling and drift velocity of the aggregate, as functions of the aspect ratio $\alpha = D_{\rm L}/D_{\rm S}$ and the Galileo number $Ga = D_{\rm L} u_{\rm ref}/\nu$. For small values of the Galileo number, the orientation of the aggregate and its velocity are seen to converge to their terminal values in a monotonic fashion, whereas, for larger Galileo numbers, the aggregate tends to behave as an underdamped oscillator. The largest drift velocities are generally observed when the aggregate is tilted at about $\pi/4$ radians with respect to the vertical direction for lower $Ga$, while for higher $Ga$ the largest drift velocities tend to be for orientations closer to horizontal. For large aspect ratios and small Galileo numbers, the terminal orientation of the aggregate tends to be vertical, whereas for smaller aspect ratios and larger Galileo numbers, the terminal orientation is inclined with respect to the vertical, which also results in a nonzero terminal drift velocity. When the Galileo number is sufficiently large for flow separation to occur, aggregates with large aspect ratios orient themselves such that the smaller sphere is located at the separation line. Empirical scaling laws are obtained for the terminal settling velocity and orientation angle as functions of the aspect ratio and Galileo number. An analysis of the accompanying fluid velocity field indicates the formation of a horizontal pressure gradient across the larger sphere, which represents the main reason for the emergence of the drift velocity, and it shows the formation of vortical structures exhibiting complex topologies in the aggregate's wake. We also note that due to these asymmetries in the aggregate, during sedimentation, the particles will likely disperse greatly from their initial locations over long settling times, indicating the importance of characterizing the relationship that the asymmetry, represented by $\alpha$, has on the velocity. 

Further work should extend the results of the present study to higher ranges of $Ga$ than those considered here. However, based on the present results we can make some predictions as to the expected behavior under those conditions. First, for higher $Ga$ one can expect $\theta_{\rm term}$ to eventually cease to converge to a steady value, as a result of vortex shedding and turbulent wake dynamics, so that we would expect periodic or chaotic behavior. Similarly, one can expect the purely planar behavior, where the aggregate remains in the $(y,z)$-plane, to also end, with the aggregate rotating or tumbling out of plane as a result of the asymmetries caused by turbulent flow past the larger particle. Additional work is required to quantify the dynamics under such conditions.

The present computational framework for establishing rigid bonds between spherical particles will enable us to explore more complex aggregates consisting of many primary particles of potentially different sizes (and potentially in different flow geometry). Simulations of this type, in turn, will allow us to investigate the influence of the porosity and permeability of the aggregate on its transient and terminal settling dynamics, and to assess how its effective settling and drift velocities are related to its geometrical properties, such as the fractal dimension. Efforts in this direction are currently underway.

\vspace{.2in}
\noindent
{\bf Acknowledgments}

This work is supported by grant W912HZ22C0037 from the Engineer Research \& Development Center at the U.S. Army Corps of Engineers, by grant W911NF-23-2-0046 from the Army Research Office, and by National Science Foundation grant CBET-2138583. The authors gratefully acknowledge several stimulating discussions with Jarrell Smith, Kelsey Fall, and Danielle Tarpley, as well as with Bernhard Vowinckel, Alexandre Leonelli, and Ram Sharma. 

\vspace{.2in}
\noindent
{\bf Declaration of Interests}\\
The authors report no conflict of interest.

\bibliographystyle{jfm}

\bibliography{draft_bib}

\clearpage

%%% SM1

\section*{Supplementary Information 1: single settling sphere validation}

\begin{figure}
    \centering
    \subfigure[]{\includegraphics[width=0.475\textwidth]{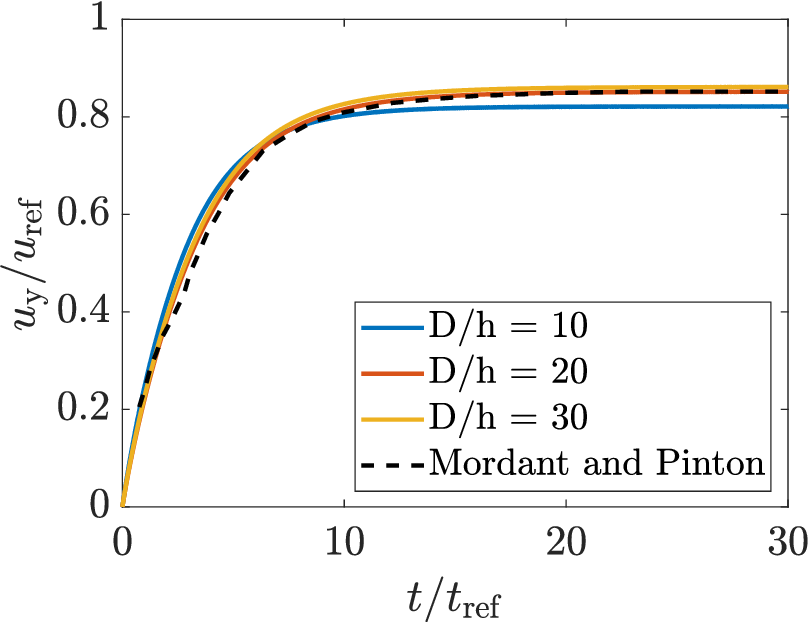}}
    \subfigure[]{\includegraphics[width=0.475\textwidth]{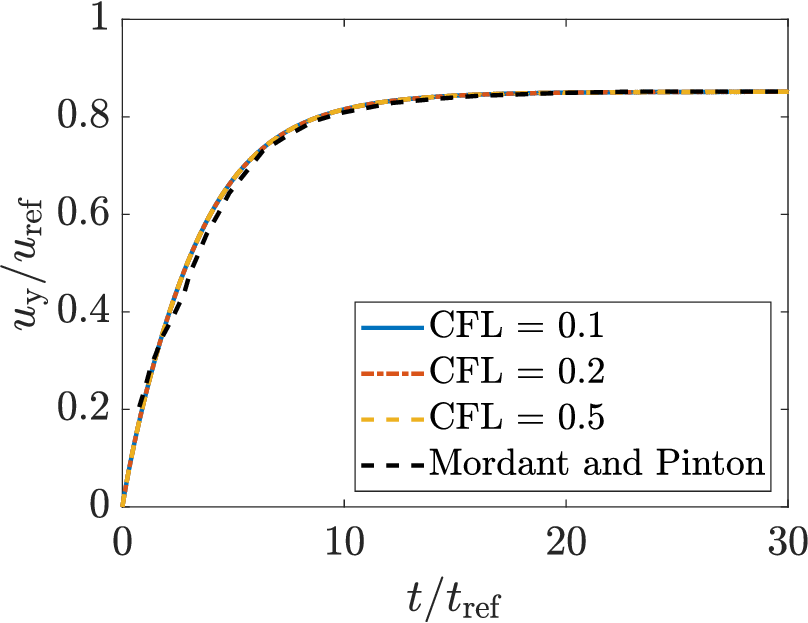}}
    \caption{Validation results for the settling velocity of a single sphere, simulated for (a) ${\rm CFL} = u\Delta t/h = 0.1$ and varying spatial step sizes $h$, and (b) spatial discretization $D/h = 20$ and varying CFL numbers. The simulation results are compared to the experimental data of \citet[][]{Mordant2000VelocityMO} for $Ga = 49.26$ and $\rho ' = 2.56$.}
    \label{fig:dx}
\end{figure}

To confirm the accuracy of our numerical method, and to estimate the spatial grid size required for convergence, we include here the case of a single particle settling in fluid at rest. We consider an experimental case from the literature, of a spherical particle settling in water \citep[][]{Mordant2000VelocityMO}. We choose to compare our numerical model against their experiment $1$, with a particle diameter $D=5\times10^{-4}\,{\rm m}$, kinematic viscosity $\nu = 0.89\times 10^{-6}\,{\rm m^2 \, s^{-1}}$, particle density $\rho_{\rm p} = 2560\,{\rm kg \, m^{-3}}$, and fluid density of water at $25^{o}{\rm C}$, $\rho_{\rm f} = 997.05\,{\rm kg \, m^{-3}}$. In our notation this corresponds to a sphere settling at $Ga = 49.26$, which is on the same order as the values of the Galileo number considered in the present experiments. We vary both the grid resolution and the time step, to assess their influence on the settling velocity.

Figure \ref{fig:dx}(a) shows that we find a good quantitative agreement between the numerical values and the experimental values obtained for the particle velocity. We also find that increasing the resolution from $D/h=10$ to $D/h=20$ improves the accuracy of the velocity obtained in the numerical model. However, there does not appear to be a significant increase in accuracy if the grid is refined beyond $D/h = 20$.

To determine the fluid time step size $\Delta t$, as a stability condition, we define the Courant–Friedrichs–Lewy (CFL) number
\begin{equation}
{\rm CFL} = \frac{u \, \Delta t}{h} \ ,
\end{equation}
where $u$ indicates the maximum vertical velocity value within the system at any given time step. Figure \ref{fig:dx}(b) demonstrates that CFL values below 0.5 give stable and accurate results for the single sphere case.

\begin{figure}
    \centering
    \includegraphics[width=0.5\textwidth]{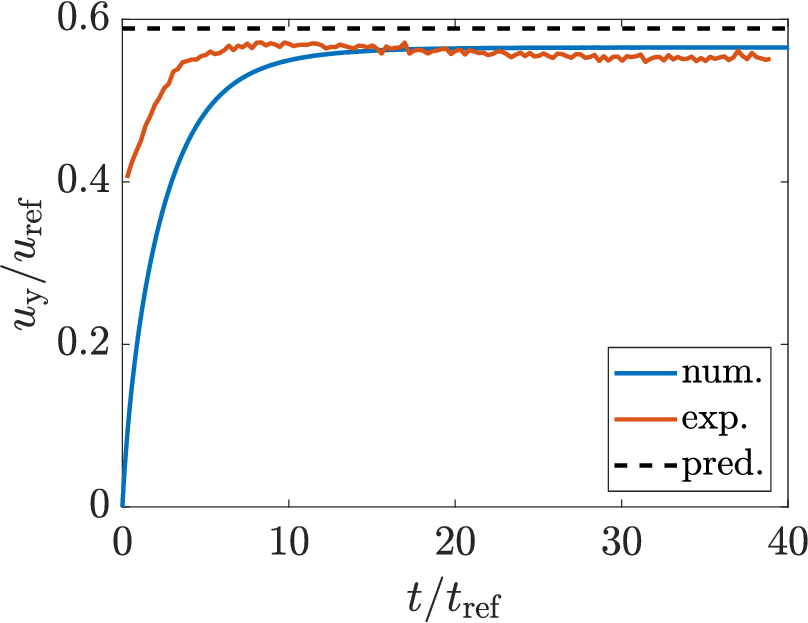}
    \caption{Comparison of experimental and simulation data for the time-dependent settling velocity of a single sphere, for $D/h=20$. The predicted velocity value shown by the horizontal dashed line is obtained from equation (\ref{eq:terminal}), for a particle of size $D = 11.1\,{\rm mm}$ and density $\rho_{\rm p} = 1135  \,{\rm kg \, m^{-3}}$ in a fluid of dynamic viscosity $\mu_{\rm f} = 0.0942\,{\rm Pa \,s}$ and density $\rho_{\rm f} = 864 \,{\rm kg \, m^{-3}}$, with $Ga = 18.81$.}
    \label{fig:single_sphere_valid}
\end{figure}

In figure \ref{fig:single_sphere_valid} we compare the results of simulations performed with $D/h=20$ for a single settling particle to our own experimental results. We also compare the results to the predicted terminal velocity $u_{\rm term}/u_{\rm ref}$, where $u_{\rm term}$ is the dimensional terminal velocity obtained via balancing the buoyancy and drag forces
\begin{equation}
u_{\rm term} = \sqrt{\frac{4g\left(\rho '-1\right)D}{3C_D}} \label{eq:terminal}  \ .
\end{equation}
Here $C_D$ is the drag coefficient as defined in equation (2.2) in the main text. The simulation results match the experimental data well with regard to the terminal settling velocity. The observed discrepancy during the initial, transient phase reflects the difficulties experimentally in starting the aggregate fully at rest.

\begin{figure}
    \centering
\includegraphics[width=0.5\textwidth]{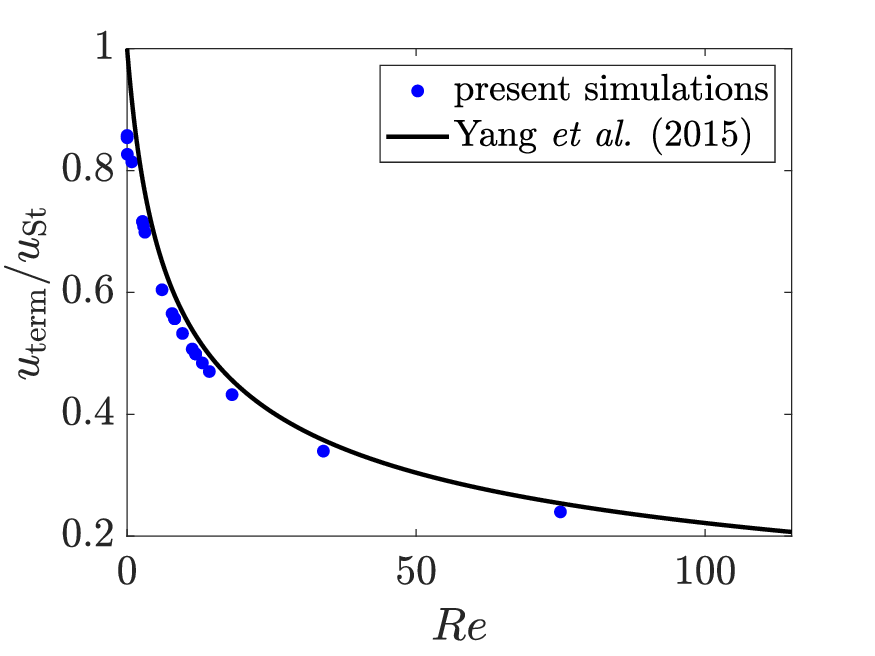}
    \caption{Terminal settling velocity of a single sphere obtained numerically compared to the corresponding data from \citet[][]{YANG2015219} (equation (42) in their paper), obtained with with $D/h = 20$ and ${\rm CFL} = 0.1$. The terminal settling velocity is normalized by the Stokes settling velocity $u_{\rm St}$. A good quantitative agreement is observed across the entire range of $Re$.}
    \label{fig:burton}
\end{figure}
Due to the potential for stability issues due to the discretization of the viscous term in the governing equations for low Reynolds (and Galileo) numbers, we additionally consider the behavior of the simulations at $\Rey < 1$. Figure \ref{fig:burton} presents a comparison of our simulation results for the terminal settling velocity with corresponding results by \citet[][]{YANG2015219} (equations (33)-(42) in their paper), across the range $\Rey \in [0.01,75]$ of interest here. In the figure the settling velocity is normalized by the Stokes settling velocity
\begin{equation}
    u_{\rm St} = \frac{2}{9}\frac{\left(\rho_{\rm p}-\rho_{\rm f}\right)}{\mu}g\left(\frac{D}{2}\right)^2 \ .
\end{equation}

Figure \ref{fig:burton} demonstrates that good agreement is observed across the entire range of Reynolds numbers, showing the accuracy of the numerical method used here for a single settling sphere.

\end{document}